\documentclass[aps,jcp,preprint,superscriptaddress]{revtex4-2} 

\usepackage{booktabs} 
\usepackage{array} 
\usepackage{paralist} 
\usepackage{verbatim} 
\usepackage{subfigure} 
\usepackage{mathrsfs}
\usepackage{amsmath}
\usepackage{dsfont}
\usepackage[titletoc]{appendix}

\usepackage[hidelinks]{hyperref} 
\usepackage{braket} 
\usepackage{subfigure} %
\usepackage{subcaption}
\usepackage{graphicx}
\usepackage{placeins} 
\usepackage[nottoc,notlot,notlof]{tocbibind} 
\usepackage{soul} 

\usepackage{geometry}                
\usepackage{graphicx}
\usepackage{amssymb}
\usepackage{epstopdf}
\usepackage{tikz}


\begin{document}

\title{A brief introduction to the diffusion Monte Carlo method and the fixed-node approximation}

\author{Alfonso Annarelli}
\email{alfonso.annarelli@gmail.com}
\affiliation{Dipartimento di Fisica Ettore Pancini, Università di Napoli Federico II, Monte S. Angelo, I-80126 Napoli, Italy}

\author{Dario Alf\`e}
\email{dario.alfe@unina.it}
\affiliation{Dipartimento di Fisica Ettore Pancini, Università di Napoli Federico II, Monte S. Angelo, I-80126 Napoli, Italy}
\affiliation{Department of Earth Sciences, University College London, Gower Street, London WC1E 6BT, United Kingdom}
\affiliation{Thomas Young Centre and London Centre for Nanotechnology, 17-19 Gordon Street, London WC1H 0AH, United Kingdom}

\author{Andrea Zen}
\email{andrea.zen@unina.it}
\affiliation{Dipartimento di Fisica Ettore Pancini, Università di Napoli Federico II, Monte S. Angelo, I-80126 Napoli, Italy}
\affiliation{Department of Earth Sciences, University College London, Gower Street, London WC1E 6BT, United Kingdom}

\date{\today}

\begin{abstract}
Quantum Monte Carlo (QMC) methods represent a powerful family of computational techniques for tackling complex quantum many-body problems and performing calculations of stationary state properties.
QMC is among the most accurate and powerful approaches to the study of electronic structure, but its application is often hindered by a steep learning curve, hence it is rarely addressed in undergraduate and postgraduate classes. 
This tutorial is a step towards filling this gap.
We offer an introduction to the diffusion Monte Carlo (DMC) method, which aims to solve the imaginary time Schrödinger equation through stochastic sampling of the configuration space. 
Starting from the theoretical foundations, the discussion leads naturally to the formulation of a step-by-step algorithm.
To illustrate how the method works in simplified scenarios, examples such as the harmonic oscillator and the hydrogen atom are provided.
The discussion extends to the fixed-node approximation, a crucial approach for addressing the fermionic sign problem in multi-electron systems. In particular, we examine the influence of trial wavefunction nodal surfaces on the accuracy of DMC energy by evaluating results from a non-interacting two-fermion system.
Extending the method to excited states is feasible in principle, but some additional considerations are needed, supported by practical insights.
By addressing the fundamental concepts from a hands-on perspective, we hope this tutorial will serve as a valuable guide for researchers and students approaching DMC for the first time.
\end{abstract}

\maketitle

\section*{Introduction}
\addcontentsline{toc}{section}{Introduction}
The development of modern physics, begun during the first decades of the $20^{th}$ century with the introduction of quantum Mechanics, provides a theoretical framework to describe the properties of materials in terms of its elementary constituents \cite{Dirac1930-DIRTPO}. What these $\acute{\alpha} \tau o \mu \alpha$ (átoma, “atoms") are, depends on the scale of the energy involved. For most applications relevant to materials modelling they are the nuclei and the electrons \cite{doi:10.1080/14786441308634955}, with the latter being responsible for most material properties, including their shape and appearance, strength, electrical and thermal conductivity, response to external perturbations, chemical reactions etc. \cite{cbd629db-42f2-34d2-8038-15d2b9a2d890,doi:10.1021/ja01379a006,doi:10.1080/00018735500101204}

These properties are described by the Hamiltonian of the system $\hat H = \hat{T} + \hat{V}$, where $\hat{T}$ and $\hat{V}$ are the kinetic and the potential energy operator, respectively.
Of particular importance is the full set of eigenstates $\{|\psi_n\rangle \}$ and eigenvalues $\{ E_n \}$ of the Hamiltonian, which in the non-relativistic limit are the solutions of the time independent Schr\"odinger equation:
\begin{equation}\label{eqn:0} 
   \hat{H} |\psi_n\rangle = E_n |\psi_n\rangle.
\end{equation}
Amongst all the eigenstates, the one corresponding to the lowest eigenvalue, the so called {\em ground state}, often plays a very special role. This is because although it is the state describing the system at zero absolute temperature, it usually remains a good descriptor of the properties of a material also at moderately high temperature (such as ambient) \footnote{See for example the ideal Fermi gas theory as discussed in most statistical mechanics books \cite{pathria1972statistical,huang1987statistical,schwabl2006statistical}.}.
However, apart from some special cases (e.g.\ the hydrogen atom), in general Eq.~\ref{eqn:0} is impossible to solve analytically and this remains so far one of the biggest challenge of chemistry and materials science.

The purpose of this tutorial is to introduce a practical method, i.e.\ the {\em diffusion Monte Carlo} (DMC) \cite{doi:10.1021/ie50503a021,10.1063/1.431514,4231595b-b4cd-332d-8271-38795f5857df,doi:https://doi.org/10.1002/9780470125908.ch3}, that provides a route to obtain the ground state $|\psi_0 \rangle$, albeit in a stochastic manner. 
It holds a prominent position among electronic structure methods due to its extreme accuracy, as well as an important historical role. Ceperley and Alder's pioneering QMC calculations of the electron gas \cite{PhysRevLett.45.566} serve as the foundation for common approximations to exchange-correlation functionals in Density Functional Theory (DFT) {\cite{PhysRev.136.B864,PhysRev.140.A1133,RevModPhys.71.1253}}, thereby playing a crucial role in enabling it. 
{Applications of the DMC method are many and varied, and it is challenging to report a chronological and exhaustive list of them here. In no particular order, it is worth mentioning calculations on diatomic molecules \cite{10.1063/1.443766}, graphene \cite{PhysRevLett.115.115501}, solid hydrogen \cite{PhysRevB.36.2092} and Compton scattering profiles \cite{PhysRevB.59.7907}.}
Furthermore, results from QMC simulations are increasingly being used to generate datasets for training machine learning models of interatomic potentials. 
Significant examples come from Kapil {\em et al.} \cite{Kapil2022} and Niu {\em et al.} \cite{PhysRevLett.130.076102}, where the phase diagram of monolayer nanoconfined water and high-pressure hydrogen, respectively, were predicted successfully using machine learning potentials with QMC as the reference method. This approach is particularly valid for computation of phase diagram or similar complex problems, as little inaccuracies in the evaluation of the interaction energy (say, less that the “chemical accuracy”, which is 1 kcal/mol) can lead to dramatically wrong prediction of stable phases.

Despite being very promising, Quantum Monte Carlo techniques are still niche methods due to their complexity and the fact that they cannot be used as black-box tools, making them hardly accessible to non-experts. 
Hence, this work is primarily aimed at students and beginners, but we also hope to provide a fresh new perspective on the topic for experienced users.
Compared to previous reviews, we showcase the techniques' advantages and difficulties via toy models and simplified systems where the effects of method's technicalities and implementation choices can be easily analysed. These are made even clearer by exploring our example codes, available on \textcolor{blue}{\href{https://github.com/zenandrea/QMC-tutorial}{GitHub}}. They focused on simplicity (much shorter than modern production codes) and clarity (making use of a few elementary functions) at the expense of efficiency. 
In fact we preferred the uniform sampling algorithm to the more efficient importance sampling one {\cite{PhysRevLett.45.566,10.1063/1.443766}}, as the latter requires more in-depth examination of subtle technical details, which we chose to discuss at a later stage.
In addition, the codes can be easily adjusted by the user to test the DMC algorithm on any kind of {single-particle} one-dimensional or three-dimensional radial potential, making them a versatile teaching tool.

Systematic errors affecting the DMC, e.g.\ the {\em time step} and the {\em population control bias} {\cite{10.1063/1.465195}}, are also analysed in this work, with an emphasis on practical examples and the different behaviour between ground state and excited state calculations. Above all, the {\em fixed-node approximation} {\cite{10.1063/1.432868,Foulkes2001}}, which is essential for dealing with multi-fermion systems, leads to a {\em trial node error} on the energy, whose functional dependence can be straightforwardly explained thanks to the simplicity of the toy models.

\section{The Schr\"odinger equation in imaginary time}\label{chap:shim}

Let us consider the time-dependent Schr\"odinger equation for a system with Hamiltonian $\hat H$ \cite{landau1991quantum}:
\begin{equation}\label{eqn:1} 
   i\frac{\partial |\phi(t)\rangle}{\partial t} = (\hat{H} - E_T) |\phi(t)\rangle,
\end{equation}
where $|\phi(t)\rangle$ is the wavefunction of the system at time $t$ and $E_T$ is an arbitrary energy offset, which obviously does not change the physical behaviour of the system. Throughout this tutorial we will use atomic units, whereby the mass of the electron $m_e$, its charge $e$ and the reduced Planck's constant $\hbar$ are all equal to one. Let us rewrite Eq.~\ref{eqn:1} in terms of the imaginary time $\tau = it$ \cite{Foulkes2001,10.1119/1.18168}:
\begin{equation}\label{eqn:2} 
   -\frac{\partial |\phi(\tau)\rangle}{\partial \tau} = (\hat{H} - E_T)|\phi(\tau)\rangle.
\end{equation}
If the Hamiltonian does not depend on time, then the imaginary time evolution of the wavefunction is: 
\begin{equation}\label{eqn:3} 
   |\phi(\tau)\rangle = e^{-\tau(\hat{H} - E_T)} |\phi(0)\rangle.
\end{equation}
The meaning of the exponential operator $e^{\hat{A}}$ can be clarified by its action on the set of the eigenstates of $\hat{A}$, being: 
\begin{equation}\label{eqn:4}
e^{\hat{A}} |\alpha_n\rangle =  e^{a_n} |\alpha_n\rangle,
\end{equation}
with $|\alpha_n\rangle$ the $n^{th}$ eigenvector of $\hat{A}$ and $\alpha_n$ the corresponding eigenvalue.
To represent the wavefunction, we can express it in terms of a complete basis set, such as the totality of the eigenstates of any operator, again, for example the eigenstates $| \psi_n \rangle$ of the Hamiltonian.  We can therefore write:
\begin{equation}\label{eqn:5} 
   |\phi(0)\rangle = \sum_n c_n |\psi_n\rangle,
\end{equation}
where the coefficients $c_n$ in the expansion are given by the projections of the wavefunction $| \phi(0) \rangle$ onto the corresponding eigenstate $| \psi_n \rangle$: $c_n \equiv \langle \psi_n | \phi(0)\rangle$. An equivalent way to express Eq.~\ref{eqn:5} is to note that the identity operator (i.e.\ the operator that leaves any element of the Hilbert state unchanged by its action on it) can be written as $\mathds{1} = \sum_n |\psi_n\rangle \langle \psi_n |$, where the sum runs over all eigenstates $|\psi_n\rangle$. If the operator has a continuous spectrum, such as the position operator, for example, then the sum is replaced by an integral and we would write $\mathds{1} = \int d{\bf R} |{\bf R}\rangle \langle {\bf R}|$. These expansions are usually referred to as {\em resolutions of the identity}.

If we now insert Eq.~\ref{eqn:5} into Eq.~\ref{eqn:3} we obtain:
\begin{equation}\label{eqn:6} 
   |\phi(\tau)\rangle = \sum_n c_n |\psi_n\rangle e^{-\tau(E_n - E_T)} ,
\end{equation}
where we have assumed $E_0 < E_1 \le E_2 \le \dots$. If we set $E_T = E_0$ we see that, provided $\langle \psi_0 | \phi(0)\rangle =  c_0 \ne 0$, in the limit of long imaginary time $\tau$ the wavefunction approaches the ground state of the Hamiltonian:  
\begin{equation}\label{eqn:7} 
    \lim_{\tau \rightarrow \infty} |\phi(\tau)\rangle =  c_0 |\psi_0\rangle.
\end{equation}
This seems useful, but of course $E_0$ is precisely what we are trying to calculate. 
Furthermore, how do we realise this imaginary time evolution? 

Let us consider the resolution of the identity written in terms of the eigenstates of the position operator, $\int d{\bf R^\prime} |{\bf R^\prime}\rangle \langle {\bf R^\prime}|$,  and introduce it in the r.h.s. of Eq.~\ref{eqn:3} between the exponential operator $e^{-\tau(\hat{H} - E_T)}$ and the wavefunction $|\phi(0)\rangle$. If we also project the wavefunction $| \phi(0) \rangle$ onto $\langle {\bf R}|$, which provides a representation of the wavefunction in terms of a specific eigenstate of the position operator, or its {\em real space representation}, we obtain:
\begin{equation}\label{eqn:8} 
   \phi({\bf R},\tau) =  \int d{\bf R^\prime}  G({\bf R},{\bf R^\prime},\tau) \phi({\bf R^\prime},0),
\end{equation}
where we have written $ \phi({\bf R},\tau) \equiv \langle {\bf R}|\phi(\tau)\rangle $. The quantity $G({\bf R},{\bf R^\prime},\tau)\equiv \langle {\bf R}| e^{-\tau(\hat{H} - E_T)} |{\bf R^\prime}\rangle$ is called {\em Green function} \cite{10.1063/1.4822960,Luchow}, and Eq.~\ref{eqn:8} provides a recipe to evolve the wavefunction $\phi({\bf R},\tau)$ with imaginary time. We see that it depends on the value of the wavefunction at a previous time {\em everywhere} in space --i.e.\ on the projection of the wavefunction onto {\em every} eigenstate of the position operator--, and on the value of the Green function. 
If we cannot {exactly} solve the Schr\"odinger equation we cannot {exactly} solve Eq.~\ref{eqn:8} either, the reason being that we are in general unable to obtain an {analytic} expression of $G({\bf R},{\bf R^\prime},\tau)$. 

There are some special cases, however, where $G({\bf R},{\bf R^\prime},\tau)$ is known analytically. One such case is that of the Hamiltonian without the kinetic term. The corresponding real space representation of the imaginary time Schr\"odinger equation becomes identical to a rate equation, for example with the wavefunction $ \phi({\bf R},\tau)$ representing the density of a population of bacteria in some {aqueous} medium at position ${\bf R}$ and time $\tau$.
The Green function for this process is:
\begin{equation}\label{eqn:10.1} 
  G({\bf R},{\bf R^\prime},\tau)  \equiv   \langle {\bf R}| e^{-\tau(\hat{V}-E_T)} |{\bf R^\prime}\rangle = e^{-\tau(V({\bf R^\prime})-E_T)}\delta({\bf R}-{\bf R^\prime}),
\end{equation}
where we have used $\langle {\bf R}| {\bf R^\prime}\rangle =  \delta({\bf R}-{\bf R^\prime})$ and $\delta$ is the delta function.  Applying this to Eq.~\ref{eqn:8} we obtain:
\begin{equation}\label{eqn:8.1} 
   \phi({\bf R},\tau) =  \int d{\bf R^\prime}  e^{-\tau(V({\bf R^\prime})-E_T)}\delta({\bf R}-{\bf R^\prime}) \phi({\bf R^\prime},0) = e^{-\tau(V({\bf R})-E_T)}\phi({\bf R},0),
\end{equation}
which shows that the process is local. Eq.~\ref{eqn:8.1} also shows that even if $\phi({\bf R},0)$ is normalised, $\phi({\bf R},\tau)$ may not be, in general. The population {of bacteria} will grow {by replication} in regions where $V({\bf R}) \le E_T$ and decrease where $V({\bf R}) \ge E_T$, and unless $E_T$ is adjusted appropriately the overall population may not remain constant. Indeed, this population fluctuation provides a feedback mechanism to adjust $E_T$ if one insists on normalisation.

\begin{figure}[htbp]
\begin{center}
\includegraphics[width=8cm]{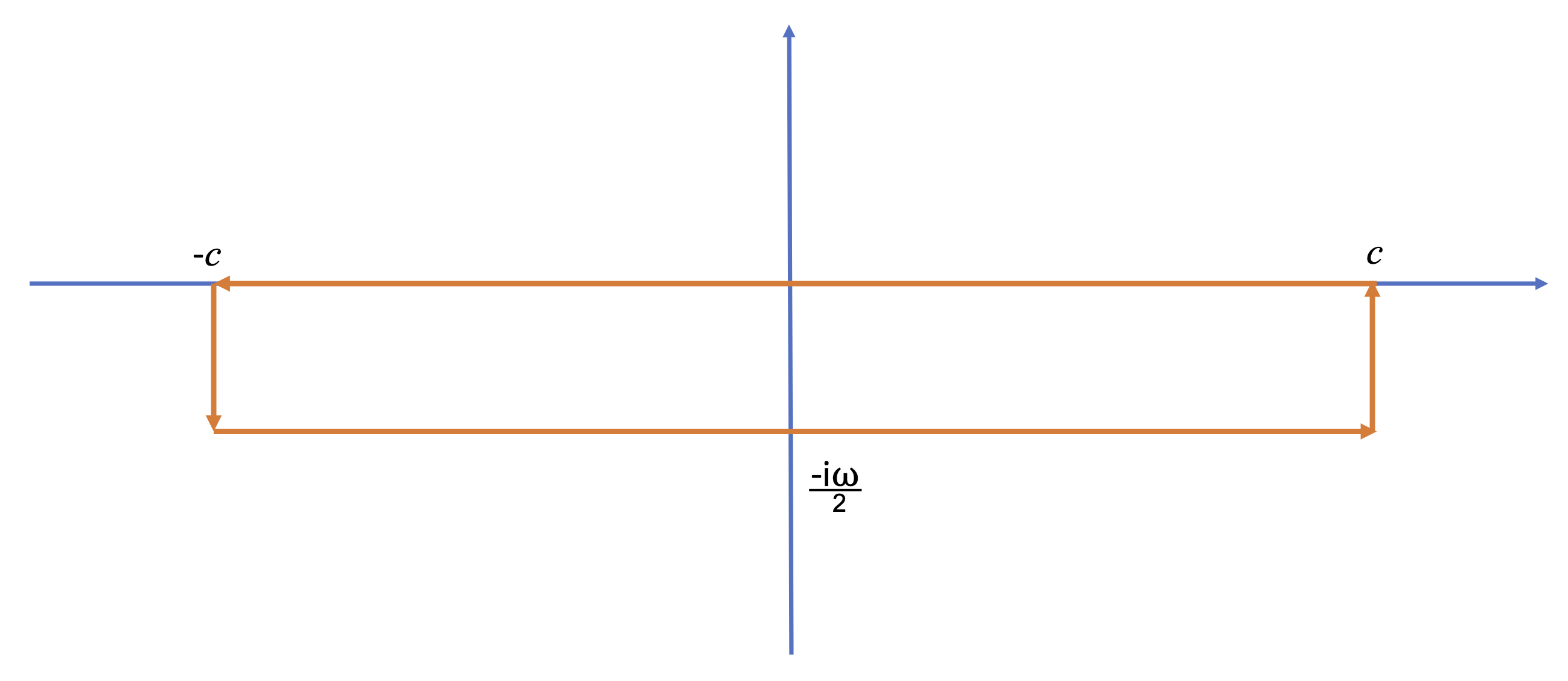}
\caption{\small The closed loop in the complex plane to evaluate the integral I.}
\label{fig:1}
\end{center}
\end{figure}

A second case for which the Green function can be obtained analytically is that for which the Hamiltonian only contains the kinetic operator,  $\hat{T} = -\frac{1}{2}\nabla^2$, and so the Green function is:
\begin{equation}\label{eqn:10} 
  G({\bf R},{\bf R^\prime},\tau)  =   \langle {\bf R}| e^{\tau\frac{1}{2}\nabla^2} |{\bf R^\prime}\rangle.
\end{equation}
The real space representation of the imaginary time  Schr\"odinger equation in this case is identical to a diffusion equation, for example describing the diffusion of bacteria in some {aqueous} medium, and again the wavefunction can be interpreted as the density of these diffusing bacteria at position ${\bf R}$ and time $\tau$.
To obtain the analytic expression of the Green function, it is useful to insert a resolution of the identity in terms of the eigenstates of the momentum operator $\hat{p} = -i \nabla$, so that we have $\hat{T} = \frac{1}{2}\hat{p}^2$ and:
\begin{equation}\label{eqn:11} 
  G({\bf R},{\bf R^\prime},\tau)  =   \int d{\bf p} \langle {\bf R}| e^{\tau\frac{1}{2}\nabla^2} |{\bf p} \rangle \langle {\bf p} |{\bf R^\prime}\rangle = \int d{\bf p} \langle {\bf R}| e^{-\tau\frac{1}{2}p^2} |{\bf p} \rangle \langle {\bf p} |{\bf R^\prime}\rangle.
\end{equation}
The real 3-dimensional space representation of the eigenstates of the momentum operator is \cite{messiah2014quantum}:
\begin{equation}\label{eqn:12} 
  \langle {\bf R}|{\bf p}\rangle = \frac{1}{(2\pi)^\frac{3}{2}}e^{i {\bf p} \cdot {\bf R}},
\end{equation}
and so:
\begin{equation}\label{eqn:13} 
  G({\bf R},{\bf R^\prime},\tau)  = \int d{\bf p} \frac{1}{(2\pi)^3}e^{i {\bf p} \cdot ({\bf R}-{\bf R^\prime)}} e^{-\tau\frac{1}{2}p^2}.
\end{equation}
To compute this integral \cite{byron1992mathematics}, first of all we recognise that it can be written as the product of the integrals in each cartesian direction, which are all equal. By making the change of variables $t = p_\alpha \sqrt{\tau / 2}$ and $\omega_\alpha = (\alpha - \alpha^\prime)/\sqrt{\tau / 2}$, with $\alpha=x,y$ or $z$, we have:
\begin{equation}\label{eqn:13.1} 
  G({\bf R},{\bf R^\prime},\tau)  = \left ( \frac{2}{\tau}\right )^\frac{3}{2}\chi(\omega_x)\chi(\omega_y)\chi(\omega_z),
\end{equation}
where the $\chi$'s are one-dimensional integrals of the type:
\begin{equation}\label{eqn:14} 
  \chi(\omega)  = \frac{1}{2\pi} \int_{-\infty}^{+\infty} dt e^{-t^2} e^{i t\omega },
\end{equation}
which can be written as:
\begin{flalign}\label{eqn:15} 
  \chi(\omega)  &= \frac{1}{2\pi} \int_{-\infty}^{+\infty} dt  e^{-t^2 + 2 \frac{it\omega}{2} - \frac{\omega^2}{4} + \frac{\omega^2}{4}}  \nonumber \\
& = \frac{1}{2\pi} e^{-\frac{\omega^2}{4}}\int_{-\infty}^{+\infty} dt  e^{-(t -\frac{i\omega}{2})^2}.
\end{flalign}
We now make the change of variable $x' = t - \frac{i\omega}{2}$ and the integral becomes:
\begin{flalign}\label{eqn:16} 
  \chi(\omega)  = \frac{1}{2\pi}  e^{-\frac{\omega^2}{4}} \int_{-\infty - \frac{i\omega}{2}}^{+\infty-\frac{i\omega}{2}} dx' e^{-x'^2},
\end{flalign}
which is a line integral in the complex plane. To evaluate it, let us introduce the closed loop integral (see Fig.~\ref{fig:1}):  
\begin{flalign}\label{eqn:17} 
I =  \frac{1}{2\pi} \left \{\int_{-c - \frac{i\omega}{2}}^{+c - \frac{i\omega}{2}} + \int_{+c - \frac{i\omega}{2}}^{+c} + \int_{+c}^{-c} + \int_{-c }^{-c - \frac{i\omega}{2}}\right \}  dx' e^{-x'^2} = 0,
\end{flalign}
which is equal to zero because there are no poles (divergences) of $e^{-x'^2}$ inside the loop. This is true for any value of $c$. 
If we now let $c \rightarrow \infty$ the two vertical segments in the loop give vanishing contributions to the integral, and we are therefore left with:
\begin{flalign}\label{eqn:18} 
  \chi(\omega)  = \frac{1}{2\pi}  e^{-\frac{\omega^2}{4}} \int_{-\infty - \frac{i\omega}{2}}^{+\infty-\frac{i\omega}{2}} dx'  e^{-x'^2} = \frac{1}{2\pi}  e^{-\frac{\omega^2}{4}} \int_{-\infty }^{+\infty} dx'  e^{-x'^2}  = \frac{1}{2 \sqrt{\pi}}  e^{-\frac{\omega^2}{4}}.
\end{flalign}
From equations (\ref{eqn:13.1}) and (\ref{eqn:18}) we get:
\begin{equation}\label{eqn:19} 
  G({\bf R},{\bf R^\prime},\tau)  = \left ( \frac{1}{2\pi\tau}\right )^\frac{3}{2}e^{-\frac{|{\bf R} - {\bf R^\prime}|^2}{2\tau}}.
\end{equation}
Notice that the Green function (\ref{eqn:19}) implies that the displacement $|{\bf R} - {\bf R^\prime}|$ is normally distributed, with a standard deviation $\sigma$ equal to the square root of the propagation time $\tau$: $\sigma=\sqrt{\tau}$.
Contrary to the rate case described previously, the diffusion process is not local. The number of bacteria --which we will more generally call {\em walkers} from now on-- at one place depends on how they diffuse from everywhere else. Note that the Green function is normalised, indicating that no walkers are destroyed or created in the diffusion process. 
It is also symmetric, $G({\bf R},{\bf R^\prime},\tau)  = G({\bf R}^\prime,{\bf R},\tau)$, which ensures {\em detailed balance} \cite{QMCbook_Hammond} and eventually allows the establishment of an equilibrium distribution. If this was not the case, the diffusion process would amplify density inhomogeneities, and prevent convergence to the correct distribution.

We start to see a picture developing. When both the potential and the kinetic term are present in the Hamiltonian they will both make their effect felt. Initially, the density of walkers will depend on time, depending on their initial distribution. The potential term will cause an exponential growth of walkers in regions of space where the potential is low, and the diffusion process will move these walkers from where the density is large to regions where the density is low.  After some equilibration time,  if the energy offset $E_T$ is appropriately chosen so that overall the number of walkers is on average constant, these two processes start to balance each other and a dynamic equilibrium is established, the density of walkers becomes time independent and their distribution represents the ground state wavefunction of the Hamiltonian. 

This seems to have solved the problem, however, there is still a small obstacle in the way, which has to do with the fact that the diffusion and the rate processes are not interchangeable. That is, if we let the walkers diffuse for some time $\tau_d$ first and then we let them replicate (or die) for some other time $\tau_r$, we obtain a distribution $\phi({\bf R},\tau_d+\tau_r)$ that is in general different from what we would obtain if we swapped the order of the two processes. In the language of quantum mechanics this is expressed by noting that the kinetic and the potential operator do not commute, it matters which one comes first.   

One may think that the two operators need to be applied simultaneously, or perhaps that applying them in short succession may reduce the difference (and so the error) caused by which one is applied first. Indeed, the latter is the strategy employed in practice to address this problem. We introduce the Trotter-Suzuki approximation \cite{Suzuki1976GeneralizedTF,PhysRevA.28.3575} to express the exponential of the sum of two generic operators $\hat A$ and $\hat B$ bounded from below:
\begin{equation}\label{eqn:20} 
  e^{-\delta\tau (\hat{A} + \hat{B})} = e^{-\frac{1}{2}\delta\tau \hat{B}}e^{-\delta\tau \hat{A}}e^{-\frac{1}{2}\delta\tau\hat{B}} + {\cal O}(\delta\tau^3),
\end{equation}
which shows that in the limit of short $\delta\tau$ it doesn't matter which operator is applied first. By taking $\hat{A} = -\frac{1}{2}\nabla^2$ and $\hat{B} = \hat{V} - E_T$, we obtain \cite{Foulkes2001,QMCbook_Hammond}:
\begin{flalign}\label{eqn:21} 
  G({\bf R},{\bf R^\prime},\delta\tau)  &= \left ( \frac{1}{2\pi\delta\tau}\right )^\frac{3}{2} e^{-\frac{\delta\tau}{2}(V({\bf R}) - E_T)}e^{-\frac{|{\bf R} - {\bf R^\prime}|^2}{2\delta\tau}}e^{-\frac{\delta\tau}{2}(V({\bf R^\prime}) - E_T)} + {\cal O}(\delta\tau^3)  \nonumber \\
  & = G_d({\bf R},{\bf R^\prime},\delta\tau) G_r({\bf R},{\bf R^\prime},\delta\tau) + {\cal O}(\delta\tau^3),
\end{flalign}
where the
\begin{flalign}\label{eqn:21.d} 
  G_d({\bf R},{\bf R^\prime},\delta\tau)  = \left ( \frac{1}{2\pi\delta\tau}\right )^\frac{3}{2} e^{-\frac{|{\bf R} - {\bf R^\prime}|^2}{2\delta\tau}}
\end{flalign}
is the Green function describing the diffusion process and 
\begin{flalign}\label{eqn:21.b} 
  G_r({\bf R},{\bf R^\prime},\delta\tau)  = e^{-\delta\tau \left(\frac{V({\bf R}) + V({\bf R^\prime})}{2}  - E_T\right)}
\end{flalign}
the one describing the rate process.

The evolution of the wavefunction for a length of time $\tau$, as in Eq.~\ref{eqn:8}, can then be split into $n=\tau/\delta \tau$ steps:
\begin{flalign}\label{eqn:22} 
   \phi({\bf R},n\delta \tau) &=  \int d{\bf R^\prime} G({\bf R},{\bf R^\prime},\delta\tau)\phi({\bf R^\prime},(n-1)\delta \tau),& \nonumber \\ 
   \phi({\bf R},(n-1)\delta\tau) &=  \int d{\bf R^\prime} G({\bf R},{\bf R^\prime},\delta\tau)\phi({\bf R^\prime},(n-2)\delta \tau),&  \nonumber \\
    \dots \dots \dots \dots  \dots \dots & \dots \dots  \dots \dots \dots \dots  \dots \dots \dots \dots  \dots \dots \dots&  \nonumber \\
    \phi({\bf R},2 \delta\tau) &=  \int d{\bf R^\prime} G({\bf R},{\bf R^\prime}, \delta\tau)\phi({\bf R^\prime},\delta \tau),&  \nonumber \\
     \phi({\bf R},\delta\tau) &=  \int d{\bf R^\prime} G({\bf R},{\bf R^\prime}, \delta\tau)\phi({\bf R^\prime},0),&
\end{flalign}
which can be written in a more compact form as:
\begin{flalign}\label{eqn:23} 
   \phi({\bf R},\tau) =  \int d{\bf R}_{n-1} \int d{\bf R}_{n-2} \cdots \int d{\bf R}_1 \int d{\bf R}_0 \, G({\bf R},{\bf R}_{n-1},\delta\tau)\\ \nonumber
   G({\bf R}_{n-1},{\bf R}_{n-2},\delta\tau) \cdots G({\bf R}_2,{\bf R}_1,\delta\tau)G({\bf R}_1,{\bf R}_0,\delta\tau)\phi({\bf R}_0,0),
\end{flalign}
leading to a total error proportional to $\delta \tau^2$.

\section{Diffusion Monte Carlo}\label{chap: diffusion monte carlo}
The steps outlined in Eqs.~\ref{eqn:22} or~\ref{eqn:23} are at the basis of the {\em diffusion Monte Carlo} (DMC) method. The term {\em diffusion} indicates the diffusion process, described by the Green function $G_d$, and {\em Monte Carlo} indicates the stochastic approach to evaluate the integrals \cite{F.James_1980} in Eqs.~\ref{eqn:22} or~\ref{eqn:23} that we will now outline. 

The wavefunction $\phi({\bf R},\tau)$ represents the density of walkers at position ${\bf R}$ and time $\tau$. If there is only one walker and we use its position to sample $\phi({\bf R},0)$, if the walker is initially at ${\bf R}_0$ we have: 
\begin{flalign}\label{eqn:24} 
   \phi({\bf R},0) = \delta ({\bf R} - {\bf R}_0).
\end{flalign}
Inserting~\ref{eqn:24} into~\ref{eqn:22} we obtain:
\begin{flalign}\label{eqn:25} 
   \phi({\bf R},\delta\tau) = G({\bf R},{\bf R}_0,\delta \tau) = G_d({\bf R},{\bf R}_0,\delta \tau)G_r({\bf R},{\bf R}_0,\delta \tau).
\end{flalign}
Since $G_d$ is normalised, if there was no $G_r$ Eq.~\ref{eqn:25} would give us the probability of finding the walker at position ${\bf R}$ at time $\delta \tau$, which means that this diffusion process can be simulated by drawing a random position ${\bf R}$ extracted from the probability distribution $G_d({\bf R},{\bf R}_0,\delta \tau)$. This walker will then have a chance to be destroyed or to be replicated depending on the value of $G_r({\bf R},{\bf R}_0,\delta \tau)$. Note that $G_r$ depends on our choice of $E_T$ (high $E_T$ more replication, low $E_T$ more destruction), and so we can affect the chances of the walker dying or be replicated by adjusting $E_T$.

If we repeat this process with $N$ walkers,  each of them will evolve according to the same process, and after a time $\delta \tau$ they will have diffused to new positions randomly extracted from $G_d$ in each case, and they will be eliminated or replicated according to $G_r$. Those walkers that find themselves in regions of space where the potential $V({\bf R})$ is low will have high chances to be replicated, and conversely those exploring regions of high $V({\bf R})$ will most likely be eliminated. After this {\emph{branching}} process, the overall number of walkers will decrease if $E_T$ is too low, or viceversa it will increase if $E_T$ is too high, and so the fluctuating number of walkers provides a natural mechanism to adjust $E_T$ so that on average their number remains constant.

After a sufficiently large number of time steps, $n_{\rm equil}$, the distribution of walkers reaches a dynamic equilibrium: at every position ${\bf R}$ in space the excess replication/elimination of walkers into an infinitesimal volume $d{\bf R}$ centred at ${\bf R}$ is balanced by diffusion out of/in to the volume $d{\bf R}$. When this happens, the distribution of the walkers resembles the ground state of the Hamiltonian $\psi_0({\bf R})$, and $E_T$ becomes an estimate of the ground state energy $E_0$. Of course, since the distribution is only sampled by the position of the walkers, there will be a statistical fluctuation associated to it. Similarly, the estimate of $E_0$ based on $E_T$ will also be affected by a statistical error. This error can be made as small as required either by increasing the number of walkers or, under the assumption of ergodicity, by continuing the simulation for a sufficiently large number of time steps $n$ and estimating $E_0$ as the average $\langle E_T^j \rangle$ over the steps $j$ after the equilibration time. This estimator of the ground state energy is called the {\em growth estimator} \cite{10.1063/1.465195}.

\subsection{The algorithm}\label{sec: algorithm}
\begin{figure}[htbp]
\begin{center}
\includegraphics[width=5.5cm]{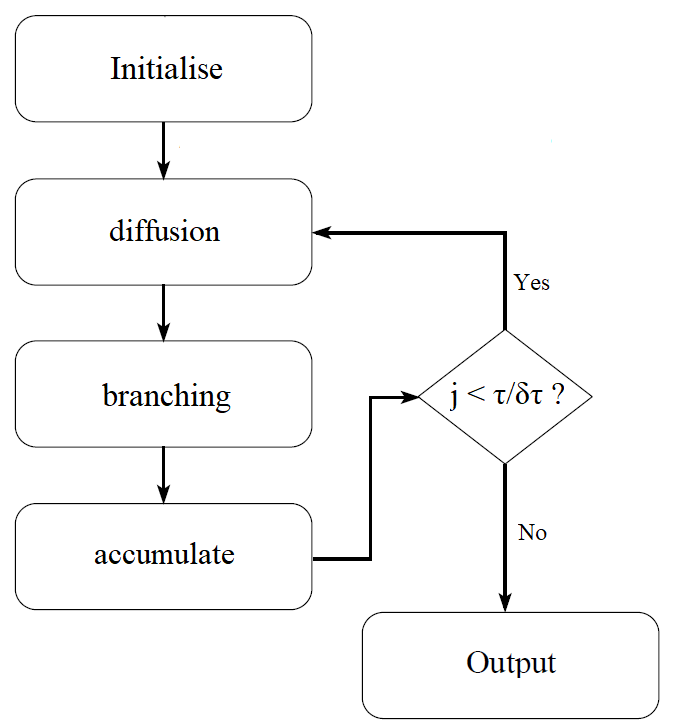}
\caption{\small Flow diagram of the DMC algorithm.}
\label{fig: DMC diagram}
\end{center}
\end{figure}

The following is a simple pseudo-code to summarise the process (see also diagram in Fig.~\ref{fig: DMC diagram}), for a three-dimensional system:

\begin{enumerate}
\item{Initialise: 
\begin{itemize}
\item{Choose a number $N_{\rm target}$ of walkers, choose $\delta \tau$, choose total number of steps $n$ of imaginary time evolution.}
\item{Distribute the $N=N_{\rm target}$ walkers at positions (${\bf R}_1,\dots, {\bf R}_N$), so that $\phi({\bf R},0)=\sum_{i=1}^N \delta ({\bf R}-{\bf R}_i)$.}
\item{Give a guess value for $E_T$.}
\end{itemize}
}
\item{Cycle over $j = 1, n$ time steps:
\begin{itemize}
\item{Cycle over the $N_j$ walkers (where $N_j$ is the number of walkers at time step $j$):}
\begin{itemize}
\item{Move ({\em diffusion}) each walker $i$ from position ${\bf R}_i$ to new position ${\bf R}_i^\prime$ with probability  $\left ( \frac{1}{2\pi\delta\tau}\right )^\frac{3}{2} e^{-\frac{|{\bf R}_i - {\bf R}_i^\prime|^2}{2\delta\tau}} $.}
\item{Evaluate $p_i = e^{-\delta\tau \left(\frac{V({\bf R}_i) + V({\bf R}_i^\prime)}{2}  - E_T\right)}$.}
\item{Evaluate the {\em branching term}: $m \equiv integer ( p_i + \eta)$, where $\eta$ is a random number uniformly distributed between 0 and 1 and {\em integer(x)} is the function returning the integer part of $x$.}
\item{if $m = 0$ eliminate the walker from the simulation, do nothing if $m=1$, otherwise add $m - 1$ copies.}  
\item{If new total number of walkers $N_{j+1}>(<)N_{\rm target}$ reduce (increase) $E_T$.}
\end{itemize}
\item{If $j > n_{\rm equil}$ {\em accumulate} ground state properties, e.g.\ the energy.}
\end{itemize}
}
\end{enumerate}

The feedback mechanism outlined in the last step has to be implemented with care, because if it is too aggressive it can cause wild fluctuations in the value of $E_T$. What is typically done is to make the update using a slow varying function of the value of $E_T$ averaged over a chosen number of previous steps $N_{\rm block}$:
\begin{flalign}\label{eqn:26}
E_T  &= E_{\rm block} - C \log \left(\frac{N_j}{N_{\rm target}}\right) \nonumber \\
E_{\rm block} &= \frac{1}{N_{\rm block}}\sum_{k=1}^{N_{\rm block}} E_T^k,
\end{flalign}
where $E_T^k$ is the value of $E_T$ evaluated $k$ steps in the past and $C$ is some appropriate energy constant that can be adjusted to modify the speed of the readjustment of the number of walkers. 

To decide how many walkers is desirable to have in a simulation it is useful to discuss one point first. The cost of moving all walkers by one time step is clearly proportional to the number of walkers $N$, but statistical sampling efficiency is also proportional to $N$, that is, if we half the number of walkers and double the number of simulation steps we still obtain the same number of samples, so this would suggest that the cost of the simulation is independent on $N$. However, only the steps $j > n_{\rm equil}$ are useful to compute ground state properties, with the first $n_{\rm equil}$ steps usually discarded. The cost of these initial $n_{\rm equil}$ steps is also proportional to $N$, and so a small $N$ would make the simulation more efficient.
On the other hand, DMC simulations are usually run on parallel computers, by distributing the walkers on different processors, and therefore the more walkers are available the more processors can be used. Load balance \cite{doi:10.1021/cr2001564} may also become an issue if there are not enough walkers on each node, as the simulation will proceed at the speed of the slowest processor, which is the one with the largest number of walkers. For this reason, walkers are re-distributed as evenly as possible after every replication/destruction, but unless they are an exact multiple of the number of processors there will be an inevitable imbalance, with some processors having $N_{\rm proc}$ walkers and others having $N_{\rm proc}+1$. For example, in the extreme limiting cases of $N_{\rm proc} = 1$ the processors with 2 walkers will run twice as slow as those who have 1 walker. Finally, the population control mechanism of Eq.~\ref{eqn:26} may bias the simulation if the number of walkers is small (more details in Sec.~\ref{sec: harmonic oscillator}).
In conclusion, we will usually aim for a large number of walkers, large enough to avoid population control biases and load balance issues, but not unnecessarily large, so to keep the equilibration time a small as possible fraction of the overall simulation.   

Another estimate of the ground state energy $E_0$ can be obtained by writing it as the expectation value of the Hamiltonian over the ground state $|\psi_0\rangle$:
\begin{flalign}\label{eqn:27} 
   E_0 & = \frac{\langle \psi_0 |\hat{H}|\psi_0\rangle}{\langle \psi_0 |\psi_0\rangle}  \\ \nonumber
& =  \lim_{\tau \rightarrow \infty} \frac{\langle e^{-\tau(\hat{H}-E_0)}\phi |\hat{H}|e^{-\tau(\hat{H}-E_0)}\phi\rangle}{\langle e^{-\tau(\hat{H}-E_0)}\phi |e^{-\tau(\hat{H}-E_0)}\phi\rangle} \\ \nonumber
 & = \lim_{\tau \rightarrow \infty} \frac{\langle e^{-2\tau(\hat{H}-E_0)}\phi |\hat{H}|\phi\rangle}{\langle e^{-2\tau(\hat{H}-E_0)}\phi |\phi\rangle},
\end{flalign}
where $\phi$ can be any function not orthogonal to $\psi_0$ (i.e.\ $\langle \phi|\psi_0\rangle \ne 0$). We choose $\phi$ to be a constant, and we replace the unknown value of $E_0$ in the r.h.s. of Eq.~\ref{eqn:27} with its estimator $E_T$. 
Of course $\phi$ cannot be exactly constant, because it would not be possible to normalise it. However, we can choose it to be constant in a sufficiently large region of space, and make it go to zero smoothly at the edges, say within a volume $V_{\rm smooth}$. There will be a kinetic energy contribution from the wavefunction in $V_{\rm smooth}$, given by $\int_{V_{\rm smooth}} -\phi({\bf R},2 \tau)\frac{1}{2}\nabla^2 \phi({\bf R}) d{\bf R}/ \int \phi({\bf R},2 \tau) d{\bf R}$, but by taking the edges sufficiently far this contribution can be made as small as wanted, as both $\phi({\bf R})$ and $\phi({\bf R},2 \tau)$ must decrease to zero at large distances.
A real space representation of Eq.~\ref{eqn:27} reads:
\begin{flalign}\label{eqn:28} 
 E_0 &\approx \lim_{\tau \rightarrow \infty} \frac{\int \phi({\bf R},2 \tau) V({\bf R})d{\bf R}}{\int \phi({\bf R},2 \tau)d{\bf R}} \\ \nonumber
 &  \approx \frac{1}{n - n_{\rm equil}}\sum_{j=1}^{n-n_{\rm equil}} \frac{1}{N_j}\sum_{i=1}^{N_j}V({\bf R}_{ij}) \\ \nonumber
  & = \frac{1}{n - n_{\rm equil}}\sum_{j=1}^{n-n_{\rm equil}} E_{\rm step}^j \equiv \langle E_{\rm step} \rangle 
\end{flalign}
with ${\bf R}_{ij}$~the position of walker $i$ at time step $n_{\rm equil} + j$, and where we have replaced the ensemble average over the distribution $\phi({\bf R},2 \tau)$ in the limit of long $\tau$ with the time average over the simulation (which is done accordingly to Eq.~\ref{eqn:23}), in which we have assumed that after $n_{\rm equil}$ times steps the distribution of walkers given by $\phi({\bf R}, n_{\rm equil} \delta \tau)$ has become stationary and proportional to the ground state of the Hamiltonian. The equivalence between the ensemble and the time averages obviously relies on an ergodic assumption \cite{doi:10.1073/pnas.18.3.263}.
The resulting estimator will be presented later (Sec.~\ref{sec: importance sampling}) in a more general form, referred to as the \emph{mixed estimator}.
In Eq.~\ref{eqn:28} we have introduced the average of the potential over the instantaneous distribution of walkers, $E_{\rm step}^j =\frac{1}{N_j}\sum_{i=1}^{N_j}V({\bf R}_{ij})$.
Note that the first $\approx$ sign in Eq.~\ref{eqn:28} is due to the approximations entering the evaluation of $\phi({\bf R},2 \tau)$ according to Eq.~\ref{eqn:23}, which are the short time step $\delta \tau$ approximation of Eq.~\ref{eqn:20} and the replacement of $E_0$ with $E_T$ in the branching term, and the second $\approx$ sign also includes the approximation of the evaluation of the integral as an average over the simulation, so it will be affected by a statistical error:
\begin{flalign}\label{eqn:29} 
 \sigma_{\langle E_{\rm step} \rangle}  = \sqrt{\frac{n_c}{n - n_{\rm equil}}}\sqrt{
\langle E_{\rm step}^2 \rangle - (\langle E_{\rm step} \rangle )^2
},
\end{flalign}
where $n_c$ is the effective number of steps that we typically need to wait to obtain two statistically independent samples, i.e.\ the {\em correlation length}, and it can be obtained by standard {\em re-blocking} procedures \cite{ErrorEstimates,Computer_Simulation_of_Liquids,PhysRevE.83.066706}. 
Notice that also the growth estimator $\langle E_T^j \rangle$ has a stochastic error $\sigma_{\langle E_T^j \rangle}$ that can be estimated with an equation analogous to (\ref{eqn:29}).

We will now show two simple applications of the methods described above, the harmonic oscillator and the hydrogen atom.
Codes (written in C) are available on {\href{https://github.com/zenandrea/QMC-tutorial}{GitHub}}.

\subsection{The harmonic oscillator} \label{sec: harmonic oscillator}
\begin{figure}
\begin{center}
\includegraphics[width=11.2cm]{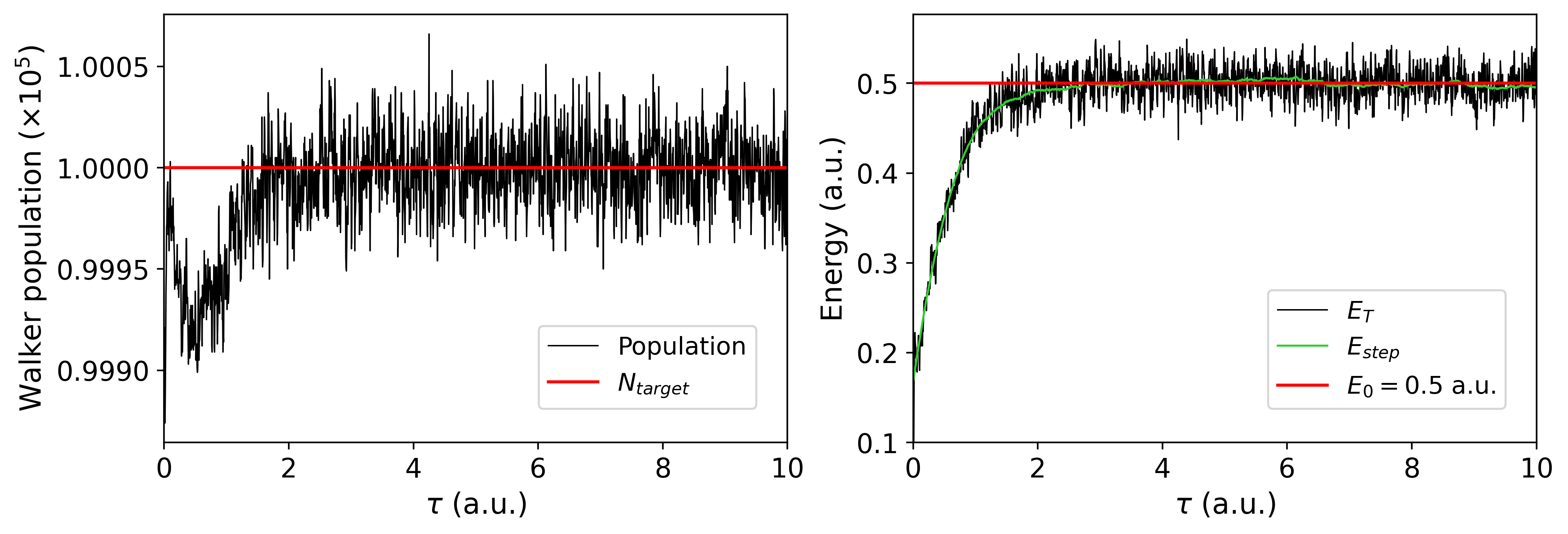}
\caption{\small Left panel: the population of walkers as function of imaginary time for the harmonic oscillator simulation described in the text. The red line is the target population chosen to be $N_{\rm target} = 10^5$. Right panel: the trial energy $E_T$ (black), the step energy $E_{\rm step}$ (green) and the exact ground state energy (red).}
\label{fig:harm1}
\end{center}
\end{figure}

\begin{figure}
\begin{center}
\includegraphics[width=7.3cm]{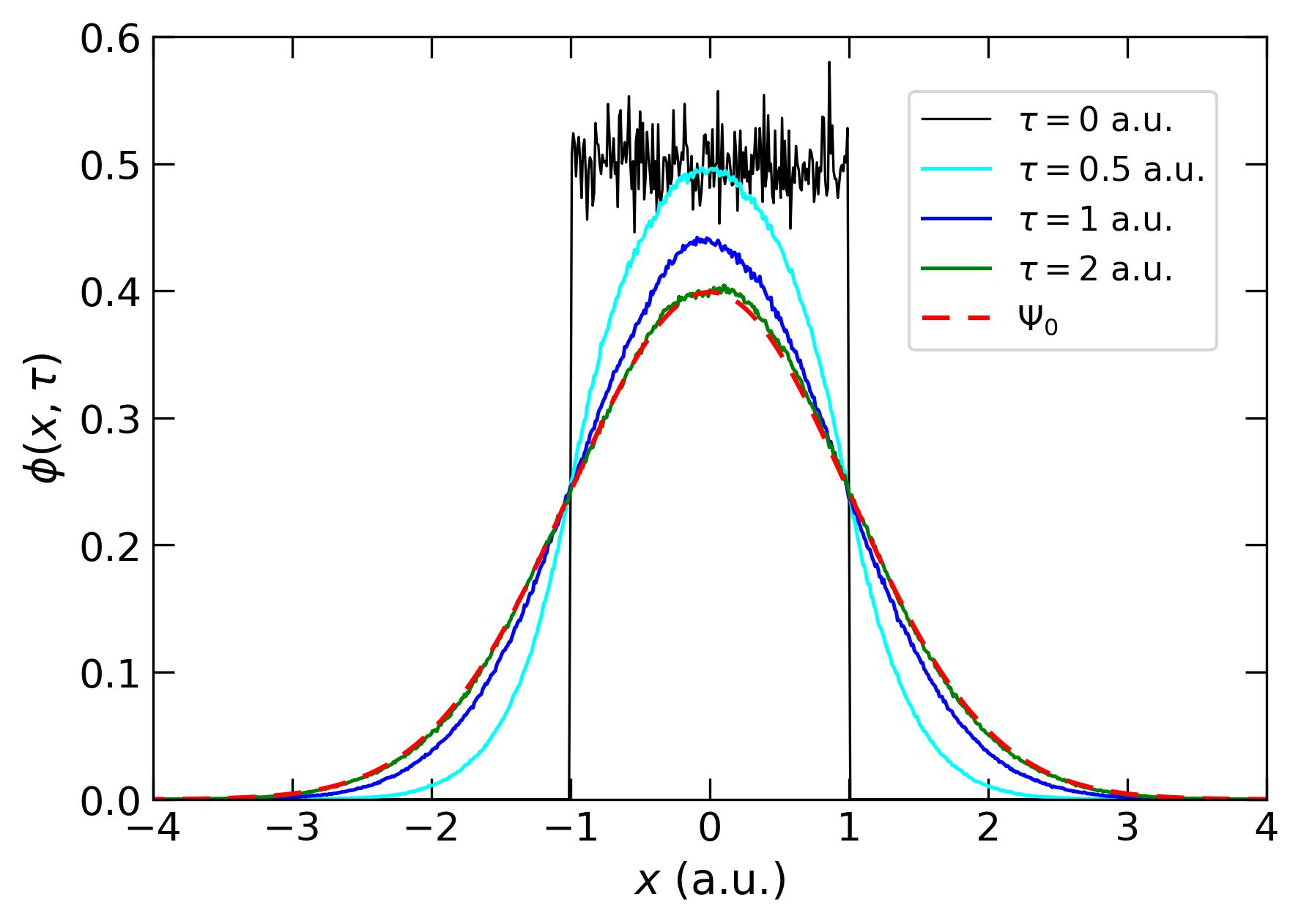}
\caption{\small The instantaneous distribution of walkers $\phi(x,\tau)$ at different imaginary times $\tau$ for the harmonic oscillator simulation described in the text, compared to the ground state wavefunction $\psi_0(x) = e^{-x^2/2}/\sqrt{2\pi}$ (here the normalisation constant has been chosen such that $\int \psi_0(x)dx=1$).}
\label{fig:harm2}
\end{center}
\end{figure}

In this section we will apply the techniques outlined in the previous sections to the one-dimensional harmonic oscillator, for which the potential energy is:
\begin{flalign}\label{eqn:30} 
 V(x) = \frac{1}{2} k x^2,
\end{flalign}
and we will chose $k =\omega^2 =1$, for simplicity. The ground state wavefunction of this system is known analytically, and apart from a normalisation constant it is equal to:
\begin{flalign}\label{eqn:31} 
 \psi_0(x) \propto e^{-\frac{x^2}{2}},
\end{flalign}
and the ground state energy is $E_0 = 0.5$ a.u.. As our first illustration, we perform a DMC simulation with $N_{\rm target} = 10^5$ and $\delta \tau =  0.005$ a.u.. For the energy constant of Eq.~\ref{eqn:26} we set $C = 1/ \delta \tau$ and we also set $N_{\rm block} = 100$. To initialise the distribution of walkers we arbitrarily choose a flat function for $ -1\le x \le 1 $ and zero outside this range (see Fig.~\ref{fig:harm2}) and we initially set $E_T=0$. Both choices are clearly significantly different from the exact ground state of the Hamiltonian. In the left panel of Fig.~\ref{fig:harm1} we show the number of walkers as function of imaginary time, and in the right panel the value of $E_T$, together with the average value of the potential energy, $E_{\rm step}$, still as function of imaginary time. The variation of the number of walkers and of $E_T$ in the initial part of the simulation depends on the choice for the initial distribution of walkers, however, we see that after $\tau \approx 2$~a.u.,  $E_T$ stabilises around 0.5 a.u., and also the population of walkers starts to oscillate around the target value.  In Fig.~\ref{fig:harm2} we show the instantaneous distribution of walkers at $\tau=0, 0.5, 1$ and $2$~a.u.. 
It changes with time, eventually becoming indistinguishable after $\tau \approx 2$~a.u.\ {from the exact distribution, which matches the ground state wavefunction $\psi_0$, as stated previously in Sec.~\ref{chap: diffusion monte carlo}. 
It is important to emphasize that this is not the general case for real system calculations, where the widely employed \emph{importance sampling} algorithm (Sec.~\ref{sec: importance sampling}) does not directly sample the ground state wavefunction, as it is biased by an arbitrary trial wavefunction selected by the user.}

\begin{figure}
\begin{center}
\includegraphics[width=7.5cm]{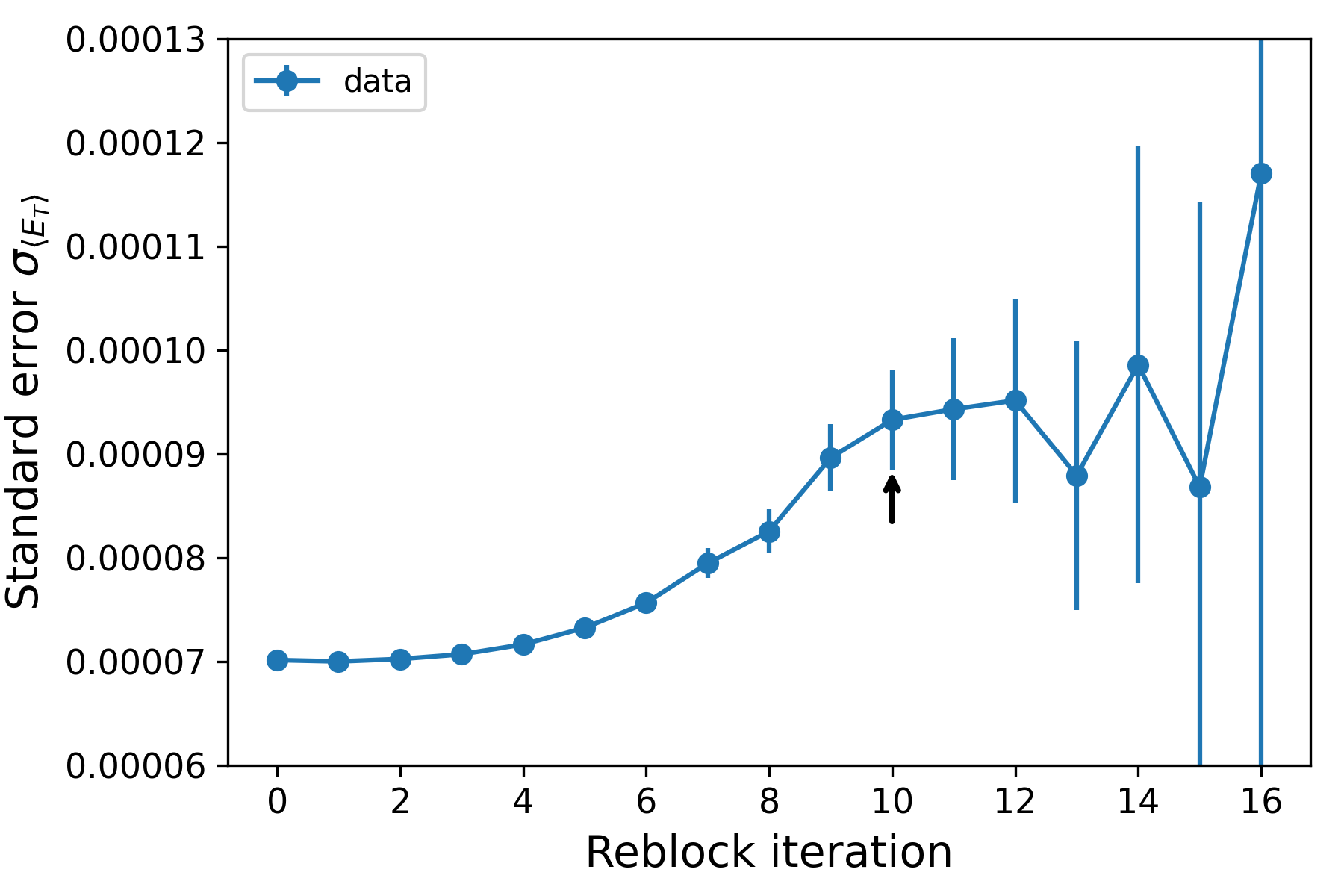}
\caption{\small $\sigma_{\langle E_T\rangle}$ (computed from block averages) as a function of the reblocking iteration number, made with the Python module \emph{pyblock}. The black arrow points out the optimal iteration number. Since the number of blocks decreases with every step, we expect an higher statistical uncertainty on the last points in the graph. DMC run performed with $\delta\tau=0.005$ a.u., $\tau=10^3$ a.u.\ and $N_{\rm target}=10^5$.}
\label{fig: reblocking}
\end{center}
\end{figure}

To obtain an estimate of the ground state energy we can now average either $E_T$ or $E_{\rm step}$ over the course of the simulation, after discarding an initial equilibration time.
Given that in a diffusion Monte Carlo simulation the configurations between successive steps often exhibit substantial {\em correlation}, the statistical error on these energy estimators must be evaluated as in Eq.~\ref{eqn:29}. The aforementioned re-blocking procedure involves partitioning the data into equal-sized blocks and generating new independent variables by averaging the measurements within each block. If the block averages are all statistically independent of each other, which is only true when $\#\textit{steps per block}>n_c$, then the estimator of the standard deviation on the average reaches the appropriate value. Plotting $\sigma_{\langle E_T \rangle}$ as a function of reblock iterations (each of which doubles the size of blocks at the previous stage) in Fig.~\ref{fig: reblocking} yields the best estimate for the standard error as the value at the plateau, while the block length at the plateau onset returns an estimate of $n_c$.

\begin{figure}
\begin{center}
\includegraphics[width=10cm]{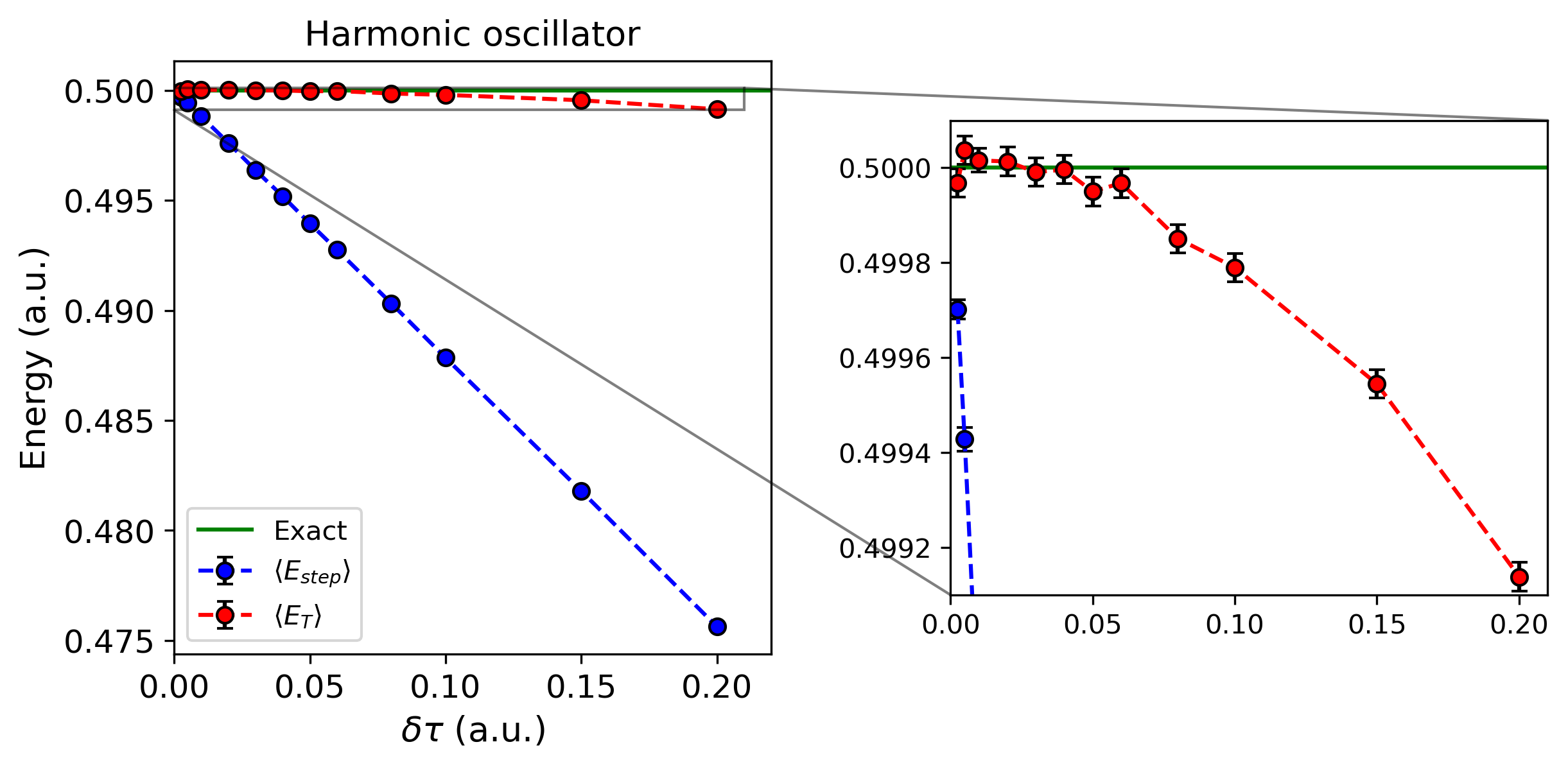}
\caption{\small The average values $\langle E_T\rangle$ (red) and $\langle E_{\rm step}\rangle$ (blue) over the course of the simulations as function of time step $\delta \tau$, for the harmonic oscillator simulations described in the text, compared to the exact value for the ground state energy of 0.5 a.u. (green line). The total length of each simulation is $10^3$ a.u. of imaginary time. }
\label{fig:harm3}
\end{center}
\end{figure}

Since the Trotter-Suzuki approximation for the Green function, Eq.~\ref{eqn:20}, is only exact in the limit $\delta \tau \rightarrow 0$, we need to perform a series of simulations with different values of $\delta \tau$, which we report in Fig.~\ref{fig:harm3}, in all cases for a total imaginary time of $10^3$~a.u.. We see that in the limit  $\delta \tau \rightarrow 0$ both energy estimators converge to the correct value $0.5$, and we note that for this system the convergence is faster for the growth estimator $\langle E_T \rangle$. 
Notice that decreasing the time step at fixed total imaginary time implies the use of a larger number of steps and then a higher computational cost, which ideally scale as $1/\delta\tau$.

\begin{figure}
\begin{center}
    \includegraphics[width=.7\textwidth, keepaspectratio]{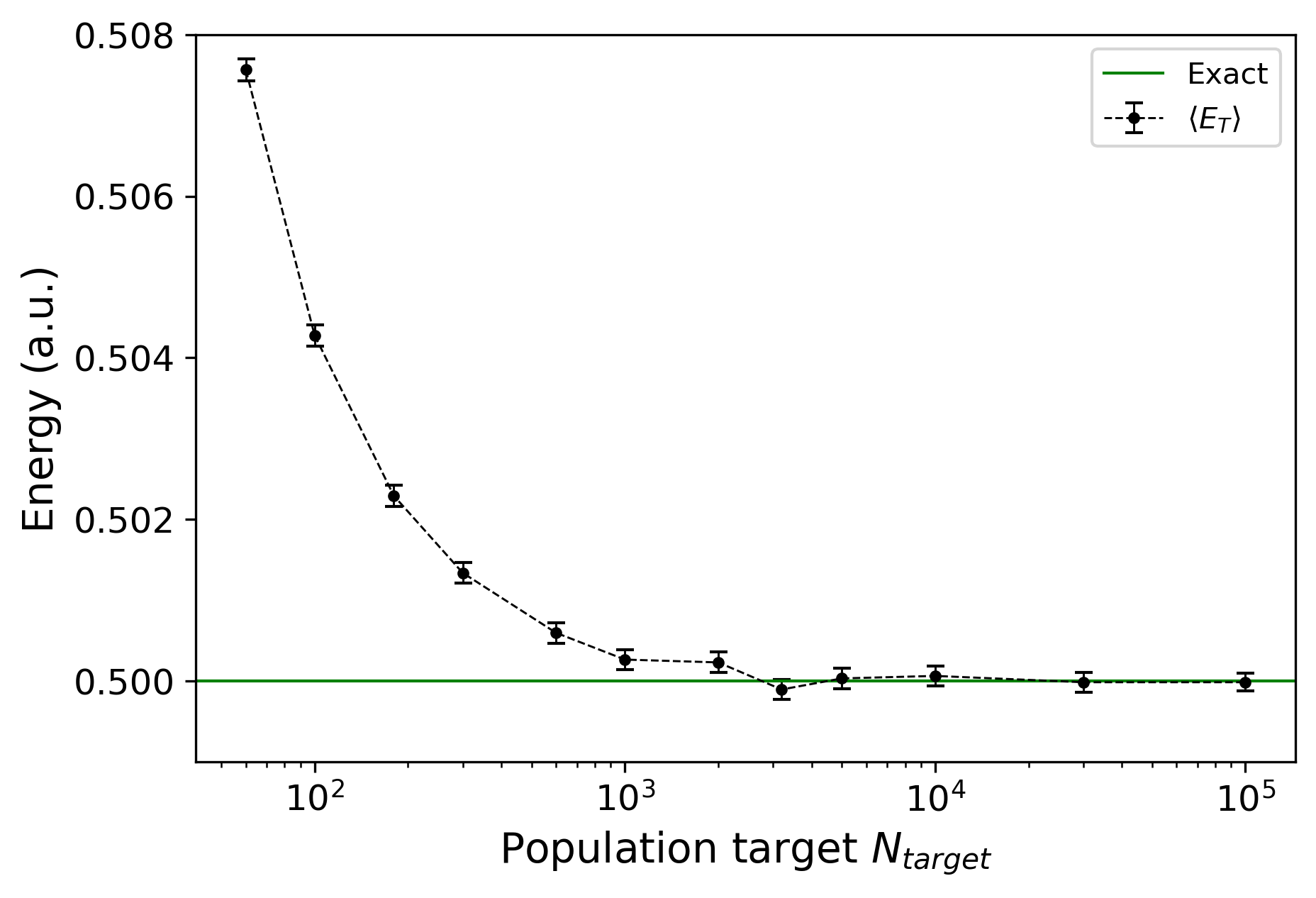}
	\caption{\small The energy estimator $\langle E_T \rangle$ as function of the target population number $N_{\rm target}$ compared to the exact harmonic ground state energy (green line). $\delta\tau=0.05$ a.u., $N_{\rm target}\cdot n=10^9$.}
    \label{fig: harm population error}
\end{center}
\end{figure}

A little residual discrepancy between the growth estimator at $\delta \tau \rightarrow 0$ and the exact value of the energy may sometimes occur as a consequence of a finite discrete population. Even though each generation of walkers on the average evolves by construction from the previous one, the feedback of the number of walkers into $E_{T}^j$ produces a \emph{population control bias} \cite{10.1063/1.465195,PhysRevB.69.085116,PhysRevB.78.125106}. This is caused by correlation of the fluctuations in the average energy of a generation $j$ of walkers and the corresponding DMC branching factor (containing $E_{T}^j$). These fluctuations, and thus the error, vanish in the ideal limit $N_{\rm target} \rightarrow \infty$. Hence, it is tested by a sequence of simulations varying the parameter $N_{\rm target}$, keeping the time step fixed. For this test, we choose $\delta \tau = 0.05$ a.u.\ because it returns a valid result for the energy, statistically equivalent to any lower time step's one. In addition, the overall number of samplings must be the same for all simulations, i.e.\ the product $(N_{target}\cdot n)$, where $n=\tau/\delta\tau$ is the number of steps, needs to be a constant, set equal to $10^9$ in the case reported in Fig.~\ref{fig: harm population error}. So the simulations with a higher population target are shorter than the others in order to maintain the same statistical accuracy on each energy estimate and consequently the same computational cost.
It's evident from Fig.~\ref{fig: harm population error} that the growth estimator converges to the exact value for $N_{\rm target}\gtrsim 3,000$, when the population control bias becomes lower than the statistical error.

\subsection{The Hydrogen atom}
\begin{figure}[htbp]
\begin{center}
\includegraphics[width=11.2cm]{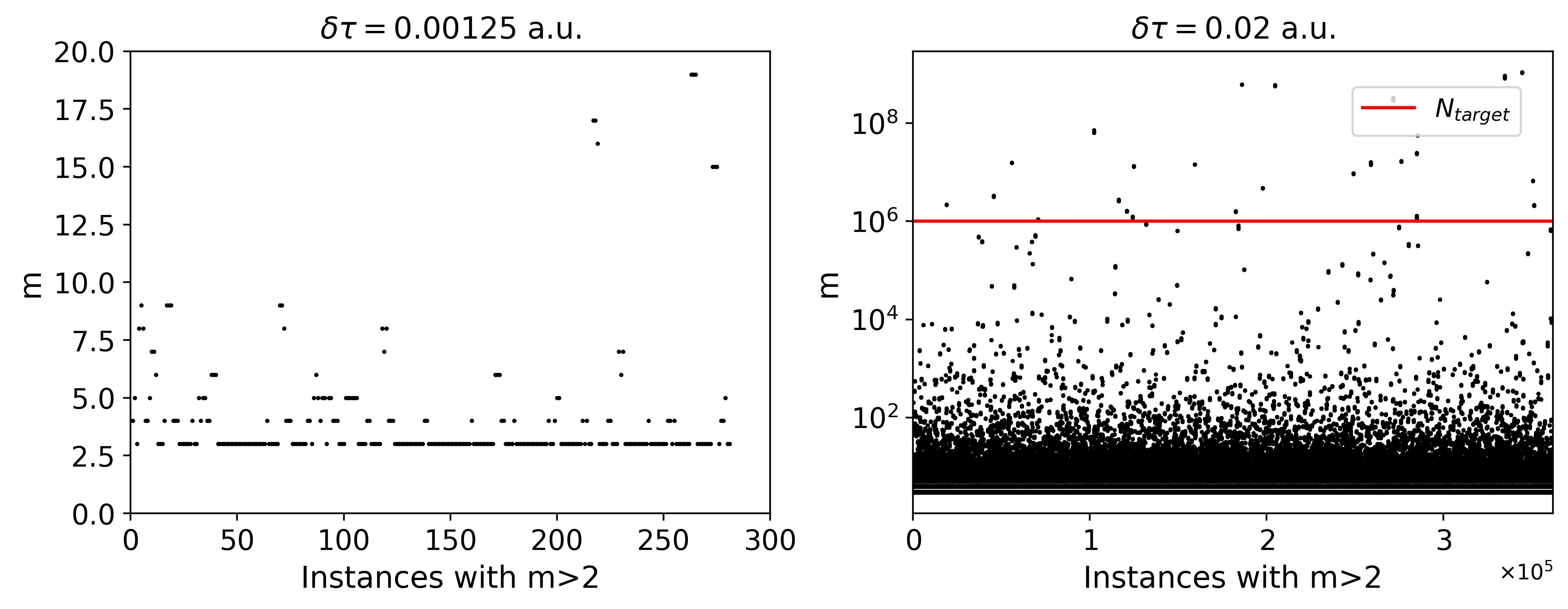}
\caption{\small The value of the branching factor $m$, plotted for instances in which it would exceed the value of 2. The left panel is for a simulation with $\delta \tau = 0.00125$ a.u. and $1.6 \times 10^6$ total time steps, and the right panel for a simulation with $\delta \tau = 0.02$ a.u. and $0.5 \times 10^6$ total time steps. The red line on the right panel shows the target population. }
\label{fig:m}
\end{center}
\end{figure}

\begin{figure}[htbp]
\begin{center}
\includegraphics[width=11cm]{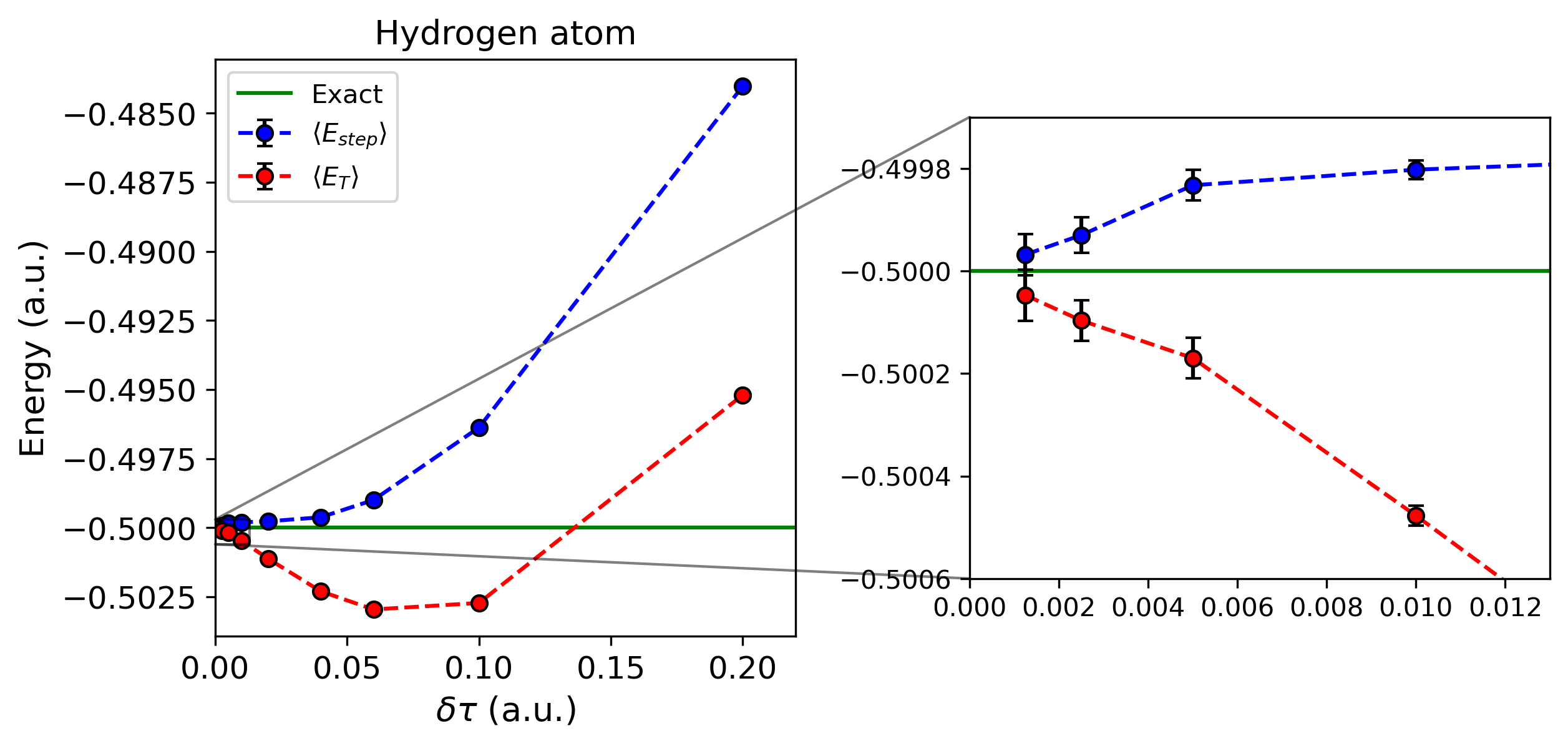}
\caption{\small The average values of the trial energy $\langle E_T \rangle$ and of $\langle E_{\rm step}\rangle$ over the course of the simulations as function of time step for the hydrogen atom, compared with the exact energy -0.5 a.u. (green line). Error bars are smaller than the size of the symbols. The graph on the right shows a zoom-in of the data for a smaller range of time steps. }
\label{fig:et_elh}
\end{center}
\end{figure}

It is useful to describe now a more realistic example of a Coulomb interaction, where the potential has a divergence (i.e.\ it corresponds to an unbounded operator), such as that of the hydrogen atom:
\begin{equation}
    V(r)=-\frac{1}{r}, \quad r=\sqrt{x^2+y^2+z^2}.
\end{equation}
This can become a problem when one or more walkers start to explore a region of space that is close enough to the divergence \cite{Martin_Reining_Ceperley_2016}. Here, for any finite value of the time step, the branching term can exceed unity by a large amount, and when this happens the Trotter-Suzuki approximation is not justified anymore. Indeed, finite time steps can take the walker accidentally very close to the nucleus, which can cause uncontrolled spikes in the population which might even be larger than the total population of walkers. These large imbalances do not typically occur very often, but when they do they may throw the simulation off-course, and make the algorithm impractical to use. One possible solution to avoid these population “explosions" is to artificially limit the value of the branching term, $m$, for example never allow it to go beyond 2 {(a similar approach has been employed, for example, in Ref.\cite{Martin_Reining_Ceperley_2016})}. This modification would bias the weight, but it would do so when the Trotter-Suzuki approximations cannot be used anyway. In the end, this bias would disappear in the limit of zero time step, because in this limit the branching term would always converge to one, {i.e.\ the Trotter-Suzuki approximation is restored,} and so this approach can be controlled by studying the simulation results in this limit. To illustrate this point, we show in Fig.~\ref{fig:m} what the value of the branching term would be if we did not limit it to 2, during the course of two simulations, one with $\delta \tau = 0.00125$~a.u. (left panel), which is run for $1.6 \times 10^6$ steps, and one with $\delta \tau = 0.02$~a.u. (right panel), which is run for $0.5 \times 10^6$ steps. In both cases we are using $10^6$ walkers. The value of the branching factor is only reported when it would exceed 2. We see that for the simulation with the short time step this happens less than 300 times, with the value of the branching factor only reaching a maximum value of 19, which would have a negligible effect on the simulation and we could also safely avoid the constraint.  However, for the simulation with the larger time step the value 2 is exceeded $3.6 \times 10^5$ times, and more importantly reaching maximum values of over $10^9$, which is more than $10^3$ times the target population. These spikes in the population would effectively be impossible to manage, and as a result a simulation with this value of the time step could not be run. 
By contrast, imposing the constraint  $m \le 2$, we can safely run simulations also with large values of $\delta \tau$, which we report in Fig.~\ref{fig:et_elh}. 
Later, we will present an improved algorithm that allows for quicker convergence in the time step without the need to impose such an artificial constraint.

\begin{figure}[htbp]
\begin{center}
\includegraphics[width=7.5cm]{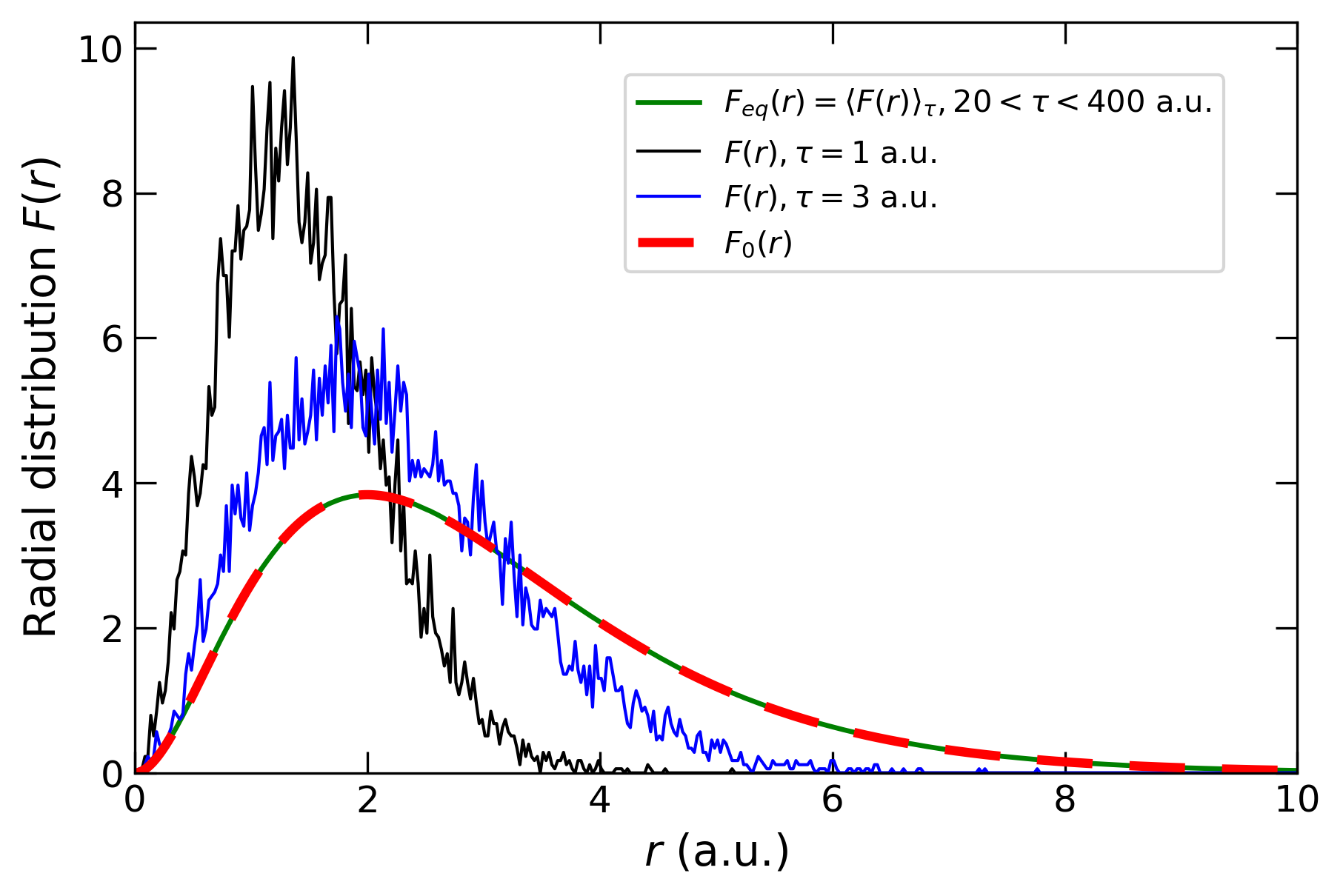}
\caption{\small Equilibrium walkers radial distribution (green solid line) compared to the exact radial distribution $F_0(r)=4\pi r^2\psi_0(r)$ of the hydrogen atom (red dashed line). Since we set $\delta\tau=0.001$ a.u.\ and $n_{\rm equil}\delta\tau=20$ a.u., the equilibrium profile has been obtained by averaging over 380,000 configurations. The instantaneous radial distributions at $\tau=1$ a.u.\ and $\tau=3$ a.u.\ are also shown as representative of the equilibration phase (black and blue solid line, respectively).
$N_{\rm target}=10^4$.}
\label{fig: atomH Phi equilibrium}
\end{center}
\end{figure}

Then we performed a new calculation with an initial rectangular distribution of walkers in all three Cartesian coordinates and a sufficiently small time step ($\delta\tau=0.001$ a.u.).
As a result, let's finally plot in Fig.~\ref{fig: atomH Phi equilibrium} the equilibrium radial distribution $F_{eq}(r) \equiv 4\pi r^2 \phi(r, \tau \rightarrow \infty)$, where $\phi(r, \tau \rightarrow \infty)$ is the projected wavefunction estimating the exact ground state wavefunction $\psi_0(r)=\exp(-r)/\sqrt{\pi}$.
To plot it, we arrange a large number of spherical bins around the nucleus position $r=0$ and then count how many walkers there are in each bin.
It is clear that with this arbitrary choice the starting distribution of walkers is very different from the ground state of the Hamiltonian, but still the algorithm converges to a suitable solution.

\FloatBarrier

\section{Fermionic systems and sign problem}
\subsection{Many particles systems}

For a many particles system the wavefunction of the ground state of any Hamiltonian is symmetric under exchange of particles. This means that the techniques described in the previous sections always result in a bosonic solution, which is characterised by a real and positive-definite wavefunction (\emph{no-node theorem} \cite{feynman1998statistical, WU_2009}). If one is interested in a system of fermions, such as any electronic system, then there is no immediate access to the corresponding lowest energy solution. That poses a significant challenge, known in the literature as the {\em sign problem}.

One possibility could be to build an anti-symmetric (fermionic) solution as a difference of semi-positive functions, say $\phi_A$ and $\phi_B$, performing two independent DMC simulations with different starting conditions, and sampling the difference $\phi = \phi_A - \phi_B$ \cite{QMCbook_Hammond}. The same arguments developed in the previous sections also apply to $\phi$, namely the fermionic eingenstates of the Hamiltonian with higher energy would decay exponentially faster than the ground state. The problem with this approach, however, is that both $\phi_A$ and $\phi_B$ converge to the bosonic ground state $\psi_0$, and so the fermionic signal in their difference, $\phi$, becomes smaller and smaller, and eventually it is overwhelmed by statistical noise. This makes it difficult, if not impossible, to accumulate statistics for the fermionic solution.
\\ {\%New paragraph\%}

A common alternative approach is to introduce what is known as the {\em fixed node approximation} (FNA) \cite{10.1063/1.432868}, whereby a {\em nodal surface} constraint is imposed, for example by introducing adsorbing walls.
The nodal surface of the function $\phi$ is the hypersurface defined by the points ${\bf R}$ where $\phi({\bf R})=0$.
A trial nodal surface, determined prior to the simulation, is generally extracted from an appropriate trial wavefunction $\psi$ which accurately approximates the ground state of the fermionic system.
In order to implement such approximation for any finite value of $\delta\tau$, a further step must be added to the usual algorithm in Sec.~\ref{sec: algorithm}: a walker moving across a node from $\mathbf{R}$ to $\mathbf{R'}$, such that $\psi(\mathbf{R})\psi(\mathbf{R'})<0$, is deleted by setting its branching term $m$ equal to 0, or alternatively the move is rejected by updating $\mathbf{R'}=\mathbf{R}$.

Of course, unless the nodal surface is identical to that of $\psi_0$, the constraint increases the energy of the system, because of the variational principle \cite{10.1063/1.443766}. 
In other words, the ground state energy estimate, which is normally a smooth function of the trial nodes position, has a minimum equal to $E_0$ and then a \emph{second-order dependence} on the nodal surface error \cite{QMCbook_Hammond,Fermion_nodes}. Let's stress that the previous considerations hold because a fermionic nodal surface (i.e.\ the one associated with an antisymmetric trial wavefunction) automatically enforces the orthogonality to every other symmetric state, including the lowest-energy nodeless ground state of $\hat{H}$.
This upper bound can be improved by improving the nodal surface with a better choice of $\psi$. The FNA usually only introduces a small error, as typically $\psi$ is taken from Hartree-Fock or Density Functional theory calculations, which in most cases do provide good quality nodal surfaces, but it is an approximation that makes DMC non exact. 

For the sake of completeness, let us now introduce some general properties related to nodal surfaces \cite{Fermion_nodes,10.1063/1.432043,KORSCH198377,10.1063/1.454227,10.1063/1.463296}. For a system with $d$ spatial dimensions and $N$ electrons, the nodal surface is formally a $(dN-1)$-dimensional hypersurface. Due to the Pauli principle, the wavefunction must vanish whenever any two electrons coincide. This defines $(dN-d)$-dimensional constraints, known as \emph{coincidence planes}, which determine completely the nodal surface only in the case $d=1$. For higher dimensionality they represent just a scaffolding through which the nodes must pass and usually no general arguments can provide further information. Certainly, the nodes should exhibit the symmetries inherent in the ground-state wavefunction. For instance, in the case of a translationally invariant Hamiltonian, the nodes must share translational invariance. It's worth noting that this imposes only a $d$-dimensional limitation on the nodes, so the constraints are not overly restrictive. \\
In addition, for the ground-state nodal surface of Hamiltonian with a local potential, a \emph{tiling property} has been proved \cite{Foulkes2001,Fermion_nodes}.
Let us first define a \emph{nodal region} or \emph{pocket}, namely a set of points in the $dN$-dimensional space that can be connected without crossing the nodes. Hence the walkers' evolution progresses independently in each region, where the algorithm returns the lowest-energy nodeless wavefunction vanishing on the boundary. In principle, a pocket $\alpha$ may yield a lower energy estimate $E_0^{\,\alpha}$ (i.e.\ {\em pocket eigenvalue}) than the others' and the gradual population correction to $E_T$ increases the walker density in the former while emptying the latter. This happens for some excited state DMC calculations, but not for the ground state ones. In fact, the aforementioned tiling theorem states that all the nodal pockets of the ground state of some local Hamiltonian $\hat{H}$ are equivalent by exchange symmetry, i.e.\ the energy estimate is the same in every nodal region.
The tiling theorem holds even when the ground state is degenerate, in which case every possible real linear combination of the degenerate ground states possesses the tiling property. It can be extended to cases where additional discrete symmetries are present. For instance, if one aims for the lowest antisymmetric state exhibiting odd parity under the inversion operator $\hat{\Pi}$, then the ground state will have the tiling property with respect to the combined action of $\hat{P}$ (particle permutation) and $\hat{\Pi}$.\\
Two main implications of the tiling theorem emerge in a fixed-node Monte Carlo calculation. Firstly, the trial wavefunction $\psi$, and thus its trial nodal surface, needs to satisfy the tiling property in order to achieve an accurate result. In general wavefunctions derived from the solution to a mean field equation, such as the local density functional approximation, are satisfactory. 
Secondly, due to the equivalence between the nodal regions, the DMC energy estimate is expected to be independent on the initial walkers distribution.

\subsection{A two-fermion system}
The toy model we are going to analyse in this section consists of two identical spinless fermions in a two-dimensional harmonic potential with angular frequencies $\omega_x=\omega_y=\omega$. Since it is a multi-particle system, its antisymmetric ground state exhibits a (three-dimensional) nodal surface, which need to be fixed a priori into the algorithm. It is described by the following non-interacting Hamiltonian (in atomic units):
\begin{equation}
  \hat{H}=\hat{H}_1+\hat{H}_2,
\end{equation}
where:
\begin{equation}
  \hat{H}_i=\frac{\hat{p}^{\,2}_{x,i}}{2}+\frac{\hat{p}^{\,2}_{y,i}}{2}+\frac{1}{2}\omega^2\hat{x}_i^{\,2}+\frac{1}{2}\omega^2\hat{y}_i^{\,2},\quad i=1,2,
\end{equation}
whose eigenstates can be factorised into the single-particle harmonic oscillator eigenfunctions:
\begin{equation}
     \psi_{n_1,m_1,n_2,m_2}({\bf r_1}, {\bf r_2})  =\psi_{n_1}(x_1)\psi_{m_1}(y_1)\psi_{n_2}(x_2)\psi_{m_2}(y_2),
\end{equation}
with eigenvalues $E_{n_1,m_1,n_2,m_2}/\omega=n_1+m_1+n_2+m_2+2$,
where
{\begin{equation*} 
\psi_{n}(x)=\frac{1}{\sqrt{2^{n}\,n!}} \left(\frac{\omega}{\pi}\right)^{1/4}\, e^{-\frac{\omega}{2}x^{2}}\,H_{n}(\sqrt{\omega}\,x), \quad n=0,1,2,..., 
\end{equation*}
and the function $H_n$ is the Hermite polynomial of degree $n$.
Since the Hamiltonian exhibits a trivial symmetry when exchanging the $x$ and $y$ coordinates of the same particle, the fermionic ground state is degenerate:
\begin{flalign}
  \psi_0^F({\bf r_1}, {\bf r_2})=\, &\alpha\,\frac{\psi_{1,0,0,0}({\bf r_1}, {\bf r_2})-\psi_{1,0,0,0}({\bf r_2}, {\bf r_1})}{\sqrt{2}}+\\ 
  &\beta\,\frac{\psi_{0,1,0,0}({\bf r_1}, {\bf r_2})-\psi_{0,1,0,0}({\bf r_2}, {\bf r_1})}{\sqrt{2}},
\end{flalign}
where the coefficients $\alpha$ and $\beta$ are arbitrary real numbers such that $|\alpha|^2+|\beta|^2=1$. \\
We chose $\omega=0.4$ a.u. and $\alpha=\beta$ for simplicity. Hence, the exact nodal surface arises from the implicit equation $\psi_0^F({\bf r_1}, {\bf r_2})=0$, namely:
\begin{equation}
    x_1-x_2+y_1-y_2=0,
\end{equation}
which is clearly invariant under particle exchange. The {\em coincidence plane}
\begin{equation}
  \begin{cases}
      x_1=x_2\\
      y_1=y_2
  \end{cases}
  \label{eqn: coincidence plane}
\end{equation}
does not provide in this case ($d=2$) enough information to identify the entire nodal surface. In order to test the behaviour of the trial node error, we can then devise an arbitrary trial nodal surface, for example:
\begin{equation}
    x_1-x_2+y_1-y_2+c\sqrt{\omega}\,(y_1^2-y_2^2)=0,
    \label{eqn: trial nodal surfaces}
\end{equation}
still passing through the coincidence plane in Eq.~\ref{eqn: coincidence plane} for every value of the real parameter $c$ (Fig.~\ref{fig: trial nodal surfaces}), and perform fixed-node calculations while varying it.
\begin{figure}[htbp]
\begin{center}
\includegraphics[width=8cm]{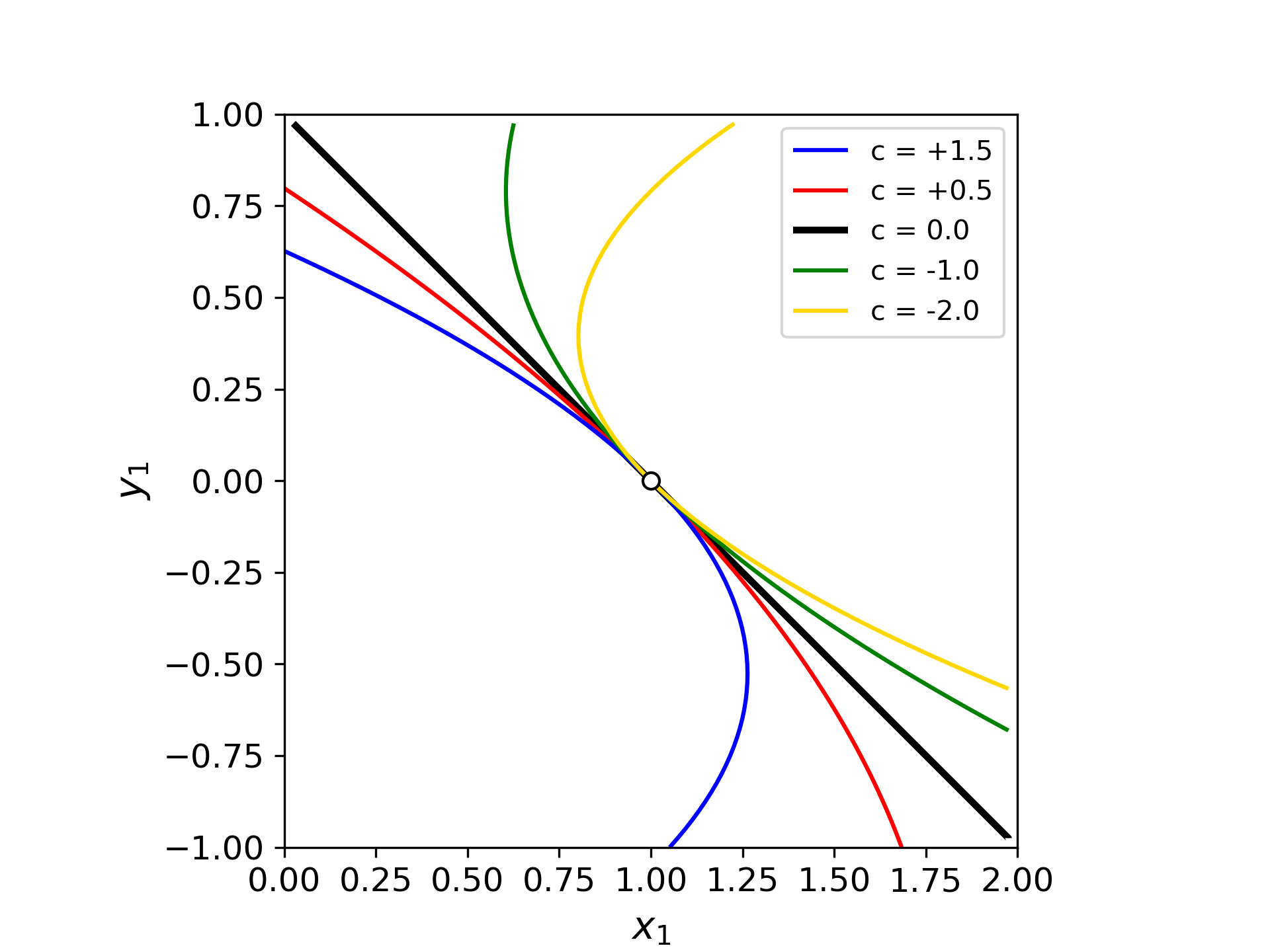}
\caption{\small A 2D cross section of the trial nodal surfaces in Eq.~\ref{eqn: trial nodal surfaces} for a two-fermion system with harmonic potential. They depend on the real parameter $c$ and the exact one is obtained for $c=0$ (black solid line). The fermion 2 is fixed at the position indicated by the open circle ($x_2=1$, $y_2=0$).}
\label{fig: trial nodal surfaces}
\end{center}
\end{figure}
They have been implemented by killing walkers that cross the nodes in a simple sampling framework, starting with a uniform walker distribution. 
Once the time step has been fixed to a sufficiently small value such that the energy bias is negligible with respect to the statistical error, the calculations return the energy profile as in Fig.~\ref{fig: harm2ferm node error}.
\begin{figure}[htbp]
\begin{center}
\includegraphics[width=11cm]{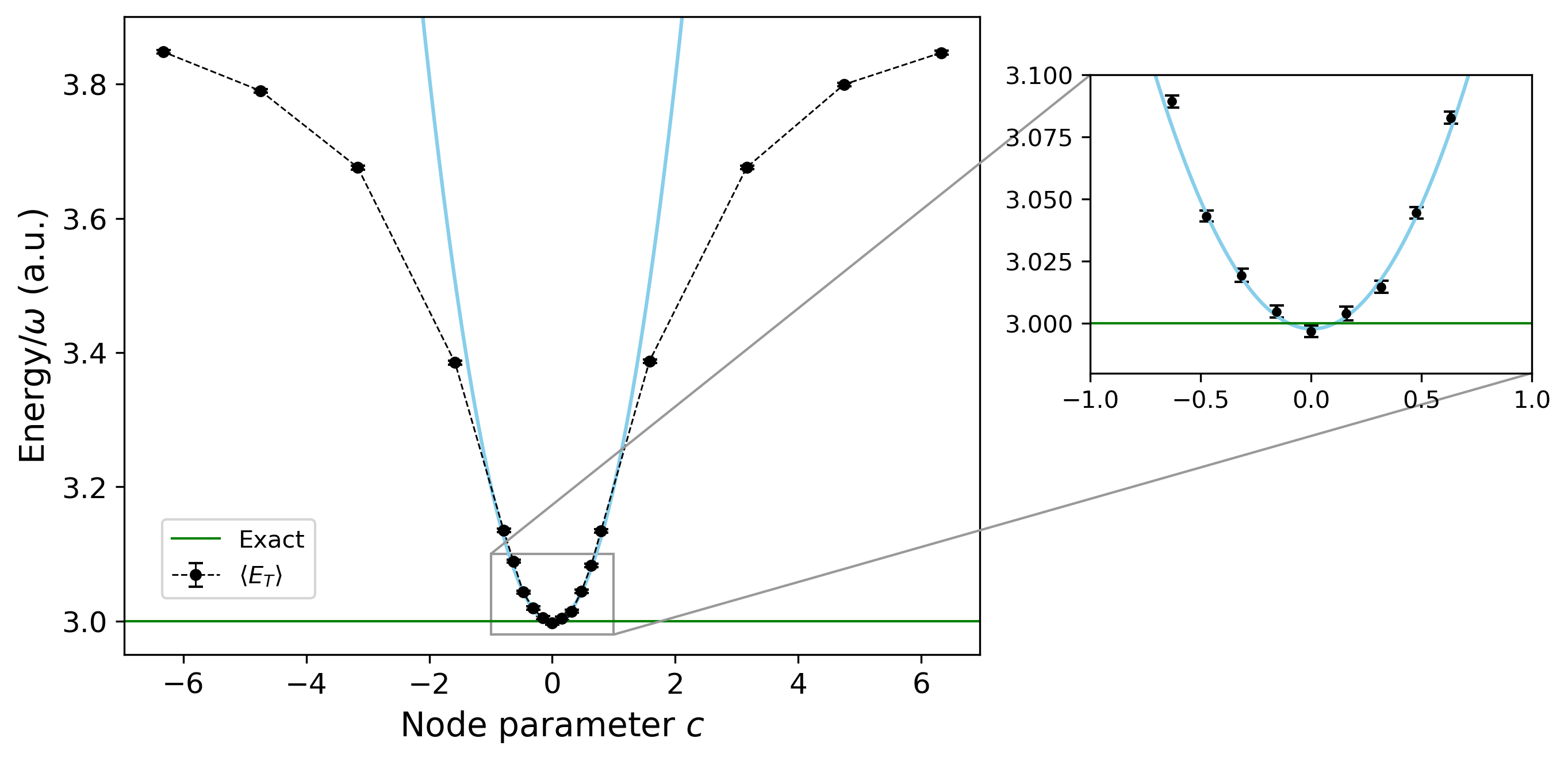}
\caption{\small DMC energy against the node parameter $c$ for a two-fermion system. The inset shows the quadratic behaviour of the energy close to the minimum at $c=0$. $\delta\tau=5\cdot 10^{-5}$ a.u., $\tau=315$ a.u., $N_{\rm target}=10^4$.}
\label{fig: harm2ferm node error}
\end{center}
\end{figure}
Although it strictly depends on the particular form of the trial nodal surfaces, the presence of a minimum at $c=0$ and the quadratic behaviour close to it clearly stand out, returning a good estimate for the ground state energy: $\langle E_T \rangle/\omega= 2.997 \pm 0.002$ a.u., to be compared with $E_0^F/\omega=3$ a.u..
Let it be noted that the trial nodal surface presented in Eq.~\ref{eqn: trial nodal surfaces} could be rejected a priori because it fails to satisfy the symmetries of the Hamiltonian (except for $c=0$), making it incompatible with a non-degenerate exact eigenstate calculation. 
Nevertheless, this does not impact the energy behavior and the example still serves the purpose of demonstrating how an incorrect trial node can influence a fixed-node DMC solution. The same applies for the trial node tested later in Sec.~\ref{sec: wrong nodes}.
\\
Despite appearing straightforward within our example, this optimization procedure for the nodal surface is rarely employed in large-scale real systems because the number of parameters increases rapidly with the system size and dimensionality and the calculations become very expensive \cite{doi:10.1021/acs.jctc.4c00139}.

\section{Excited states}
Excited states, both vibrational and electronic, are pivotal in the understanding of the physics and chemistry of atoms, molecules, and condensed matter. As discussed earlier, the fermion ground state is essentially an excited state, it represents the lowest antisymmetric state of a system and its energy is greater that the boson one. This seems to suggest that the fixed-node approximation could be generalised to the calculation of an excited state $\psi_i$ by employing a trial nodal surface from a proper trial wavefunction $\widetilde{\psi}_i$.
According to the variational principle \cite{levine2014quantum}, for the DMC solution to avoid collapsing into lower-energy states, $\widetilde{\psi}_i$ should be orthogonal to all the lower eigenstates $\psi_j$, $j<i$ {\cite{PhysRevE.84.046705,doi:https://doi.org/10.1002/9781119417774.ch8}}. Thus orthogonality can be often achieved by imposing symmetry conditions on $\widetilde{\psi}_i$ \cite{10.1063/1.451754}. Indeed, if the selected trial wavefunction has a defined symmetry that meets some specific mathematical conditions~\footnote{As stated in the Foulkes \emph{et al.}'s work \cite{PhysRevB.60.4558}, the selected trial wavefunction must transform according to a one-dimensional irreducible representation of that symmetry group.} (for example the \emph{gerade/ungerade} states in centrosymmetric molecules), the variational principle ensures that the computed energy is either equal to or greater than the eigenvalue of the lowest exact eigenstate with that particular symmetry \cite{PhysRevB.60.4558}. 
However, such considerations usually do not completely specify the exact solution.
For excited states which are energetically not the lowest in their symmetry, neither the variational principle nor the tiling property are guaranteed and one may expect a strong dependence of the result on both the trial nodal surface and the initial distribution of walkers \cite{PhysRevB.60.4558,Reynolds_Barnett_Hammond_Lester_1986,10.1063/1.454228,doi:https://doi.org/10.1002/9781119417774.ch8}.
In particular the energy estimator from Eq.~\ref{eqn:28} for the eigenstate $\psi_i$ will be expressed as a linear combination of energies that include the lower states:
\begin{equation}
  \widetilde{E}_i=\Bigl\langle V\Bigr\rangle_{\phi({\bf R},\tau\rightarrow\infty)}^{\{\widetilde{\psi}_i({\bf R})=0\}}=\sum_{j=0}^\infty c_{ij}E_j,
  \label{eq: energy mixing}
\end{equation}
where the average of the potential energy has been computed by sampling the equilibrium distribution $\phi({\bf R},\tau\rightarrow\infty)$ with nodes fixed by the implicit equation $\widetilde{\psi}_i({\bf R})=0$.
The coefficients $c_{ij}$ are related to the overlap integral $\braket{\widetilde{\psi}_{i}|\psi_j}$.
A similar expression applies also to the growth estimator.
The incorrect placement of nodes leads to a mixing of energies and therefore to a deviation from the ideal situation $c_{ij}=\delta_{ij}$.\\
In case the trial nodal surface does not possess the tiling property, the most favourable nodal pocket, namely the one with the lowest energy of all those initially occupied, becomes the only populated region in the large $\tau$ limit \cite{PhysRevB.60.4558}. In other words, the DMC energy converges to $\widetilde{E}_i= \min_\alpha\{E_i^{\,\alpha}\}$, where the index $\alpha$ runs over all the filled nodal regions. Everything introduced so far will be clarified with an example in the next section.

\subsection{First excited state of the harmonic oscillator}\label{sec: first excited state of the harmonic oscillator}
Let us investigate again the single-particle system with a 1D harmonic potential, but now we focus on the first excited state. It is described by the wavefunction:
\begin{equation}
  \psi_1(x)=\left(\frac{\omega}{\pi}\right)^{\frac{1}{4}}\sqrt{2\omega}\,x\exp \left(-\frac{1}{2}\omega x^2\right),
\end{equation}
with energy $E_1/\omega=3/2$ a.u.. We set again $\omega=0.4$ a.u.\ in the following calculations. The nodal surface of such simple system is exactly known, it is just the point $x=0$, which divide the 1D space into two equivalent nodal regions, and it needs to be enforced with the usual fixed-node constraint. 
Check the evolution of the walkers distribution along the simulation in Fig.~\ref{fig: harm1node equilibration phi}. The initial position of each walker is generated from a uniform distribution with a non-zero component in both nodal regions.  
\begin{figure}
    \begin{center}
    \includegraphics[width=11cm]{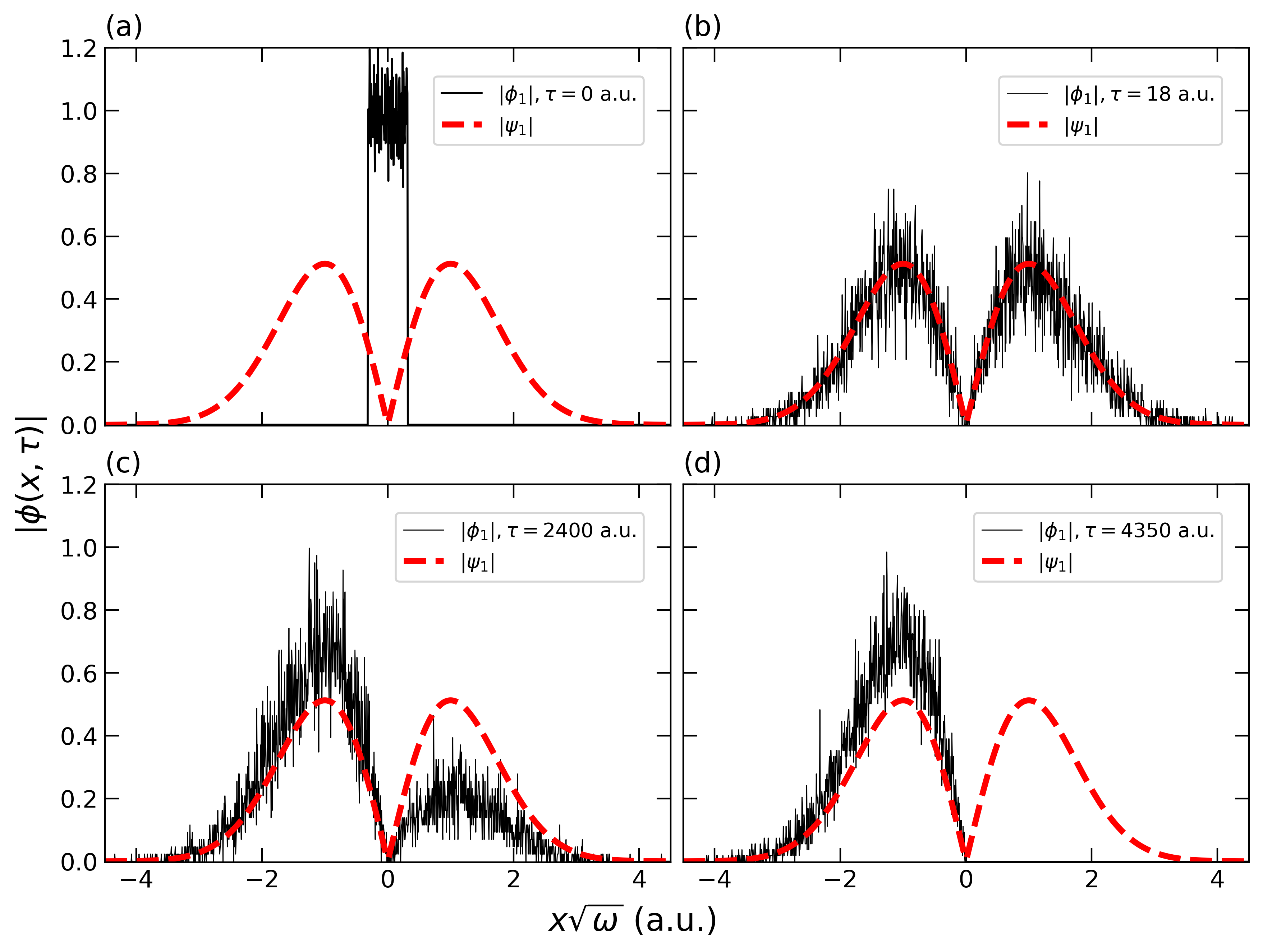}
    \caption{\small Evolution of the walkers distribution $\left|\phi_1(x)\right|$ (black line) compared to the exact first excited state distribution $\left|\psi_1(x)\right|$ (red dashed line). Simulation performed with $\delta\tau=10^{-4}$ a.u.\ and $N_{\rm target}=10^4$.}
    \label{fig: harm1node equilibration phi}
    \end{center}
\end{figure}
The population rapidly converges to the exact distribution profile (except for statistical noise, Fig.~\ref{fig: harm1node equilibration phi}(b)) but, since the node plays the role of an infinite barrier in $x=0$, the system reveals in the long $\tau$ limit a spontaneous symmetry breaking following a random fluctuation in the walkers distribution, which become more and more concentrated in just one region.
Then the $\tau \rightarrow \infty$ distribution outlines the corresponding nodal pocket eigenstate (Fig.~\ref{fig: harm1node equilibration phi}(d)), which is zero outside and still proportional inside the same nodal region to $\left|\psi_1(x)\right|$ apart from a normalization factor. Although only one pocket is sampled in the long time limit, the DMC energy comes to be accurate as long as the trial node corresponds to the exact one.

Adding a node constraint also has some important effects on the time step bias. After fixing the target population at a reasonable value (e.g.\ $N_{\rm target}\sim 10^4$, in order to have negligible population bias), a series of simulations have been performed varying $\delta\tau$ and the resulting energies are plotted in Fig.~\ref{fig: harm1node time error}, along with the analogous data for the ground state.
\begin{figure}
    \begin{center}
    \includegraphics[width=0.75\textwidth]{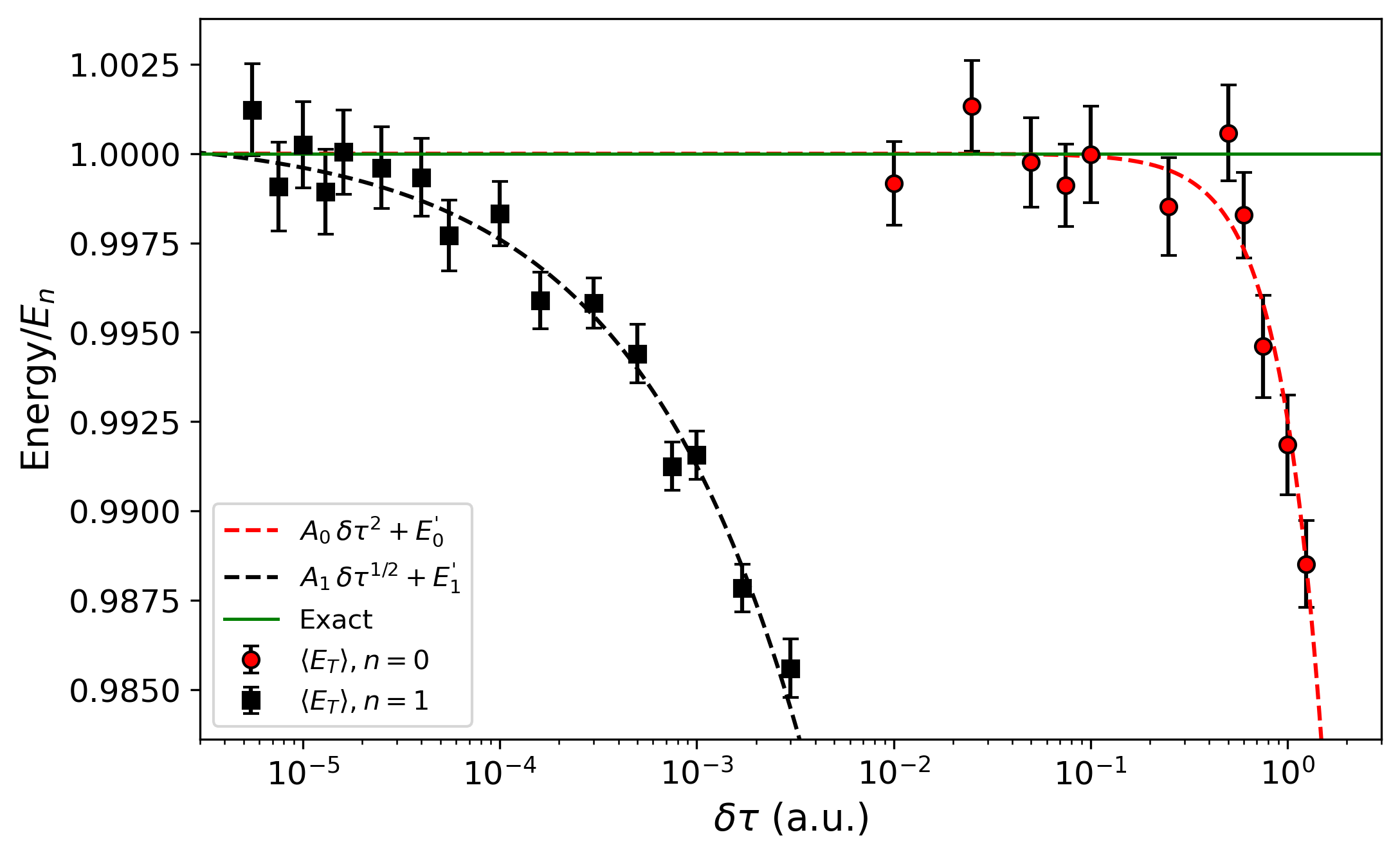}
    \caption{\small The growth estimator $\langle E_T \rangle$ as function of the time step $\delta\tau$ has been modelled according to a power law with distinct exponents for the ground state (2) and the first excited state (1/2). The energy axis has been rescaled such that both curves have the same asymptotic value at $\delta\tau=0$. Simulations performed with $\tau=410$ a.u.\ and $N_{\rm target}=10^4$.}
    \label{fig: harm1node time error}
    \end{center}
\end{figure}
\\
A slower convergent behaviour is apparent for the excited state, in particular an accurate energy estimate is achieved only with $\delta\tau=10^{-5}\div10^{-4}$ a.u., whereas for the corresponding nodeless ground state the same is attained with a way larger time step, namely $\delta\tau=10^{-2}\div10^{-1}$ a.u.. Hence, this results in expensive simulations, which are commonly optimised by employing the importance sampling method (see Sec.~\ref{sec: importance sampling}) in every practical application. Such difference in convergence efficiency can be fully understood in light of the power law behaviour of the time step bias. It can be proved \cite{PhysRevE.61.2050,PhysRevA.42.6991} that, for nodeless states and adopting the symmetric form of the Trotter-Suzuki approximation as in Eq.~\ref{eqn:20}, the trial energy converges quadratically to its exact value as $\delta\tau$ approaches $0$, i.e.\ $\langle E_{T} \rangle(\delta\tau)-E_0\sim \delta\tau^2$.
Conversely, the act of killing walkers that cross nodes in the simple DMC algorithm leads to a $\delta\tau^{1/2}$ term in the same formula and therefore to a slower convergence, as supported by qualitative arguments presented in Refs.~\cite{10.1063/1.465195,10.1063/1.443766}.
This square root dependence of the growth estimator makes accurate extrapolation to $\delta\tau=0$ difficult since $\sqrt{\delta\tau}$ has infinite slope here.\\
The fitting procedure returns the best parameters as in Table~\ref{tab: fit parameters}.
Both extrapolated energies predict correctly the exact eigenvalues with almost the same accuracy, but let's stress again that the calculation for $E_1'$ is much more expensive as it is necessary to compute many low-time step points.
\begin{table}[ht]
\centering
\caption{Fitting parameters (in units of $\omega$) for the time step bias calculation.}
\label{tab: fit parameters}
\begin{tabular}[t]{m{3em} c c}
\hline
\hline
\addlinespace
&Ground state &First excited state\\
&{\small $E (\delta\tau) =A_0\,\delta\tau^{2}+E_0'$}&{\small $E (\delta\tau) =A_1\,\delta\tau^{1/2}+E_1'$} \\
\addlinespace
\hline
\addlinespace
$A_i$&$-0.0037 \pm 0.0004$&$-0.440 \pm 0.015$\\
\addlinespace
$E_i'$&$0.5000 \pm 0.0002$&$1.5008 \pm 0.0005$\\
\addlinespace
\hline
\hline
\end{tabular}
\end{table}
\FloatBarrier
\subsection{Effects of wrong nodes on excited states}\label{sec: wrong nodes}
We show the impact of the trial node error on the growth estimator for an excited state. The 1D harmonic oscillator Hamiltonian is invariant under inversion $\hat{\Pi}$, i.e.\ $V(-x)=V(x)$, so the same must be true for the nodal surface of every excited state. Clearly, the point $x=0$ is mapped into itself by an inversion transformation and no other single point can be. In a more formal way, only a wavefunction with a single node in $x=0$ fulfills the tiling property with respect to $\hat{\Pi}$.
If we fix a node at any point $x_{node}$ other than $0$, for example $x_{node}>0$, we expect to split the 1D domain into two non-equivalent nodal regions and get a DMC energy lower than $E_1$. In particular we start the simulation with a uniform distribution of walkers centered at $x=0$ and large enough such that each nodal region is sampled during the initial phase. As proved for the $2s$ excited state of the hydrogen atom by Foulkes {\em et al.}\cite{PhysRevB.60.4558} using a variational argument, the favourable pocket is the one enclosing the exact nodal point, i.e.\ the region $x<x_{node}$ in Fig.~\ref{fig: harm1node +0.2 phi}. Indeed it is the only populated one in the high $\tau$ limit.
\begin{figure}[htbp]
    \begin{center}
    \includegraphics[width=11cm]{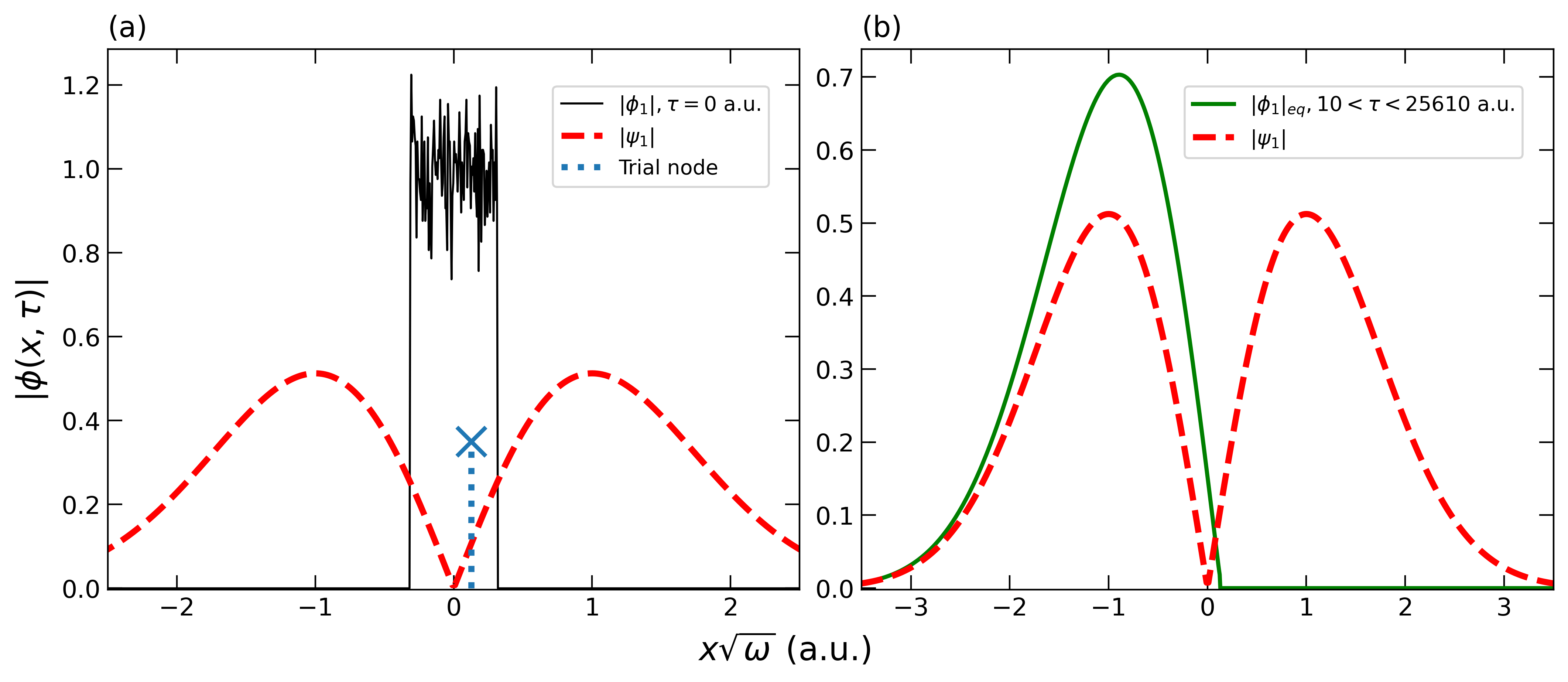}
    \caption{\small Fixing the node at $x_{node}>0$ and (a) populating both nodal regions as initial condition lead to (b) a final (averaged) distribution of walkers entirely enclosed into the most favourable pocket $\bigl\{x<x_{node}\bigr\}$. $x_{node}\cdot \sqrt{\omega}=0.2\,\sqrt{0.4}\approx 0.13$ a.u., $\delta\tau=0.001$ a.u., $N_{\rm target}=10^4$.}
    \label{fig: harm1node +0.2 phi}
    \end{center}
\end{figure}
This leads to a linear dependence of the energy on small node displacement \cite{becca_sorella_2017} and, due to the inversion symmetry of the system, to a slope discontinuity at $x_{node}=0$ (Fig.~\ref{fig: harm1node node error T410}).
\begin{figure}
    \begin{center}
    \includegraphics[width=0.78\textwidth]{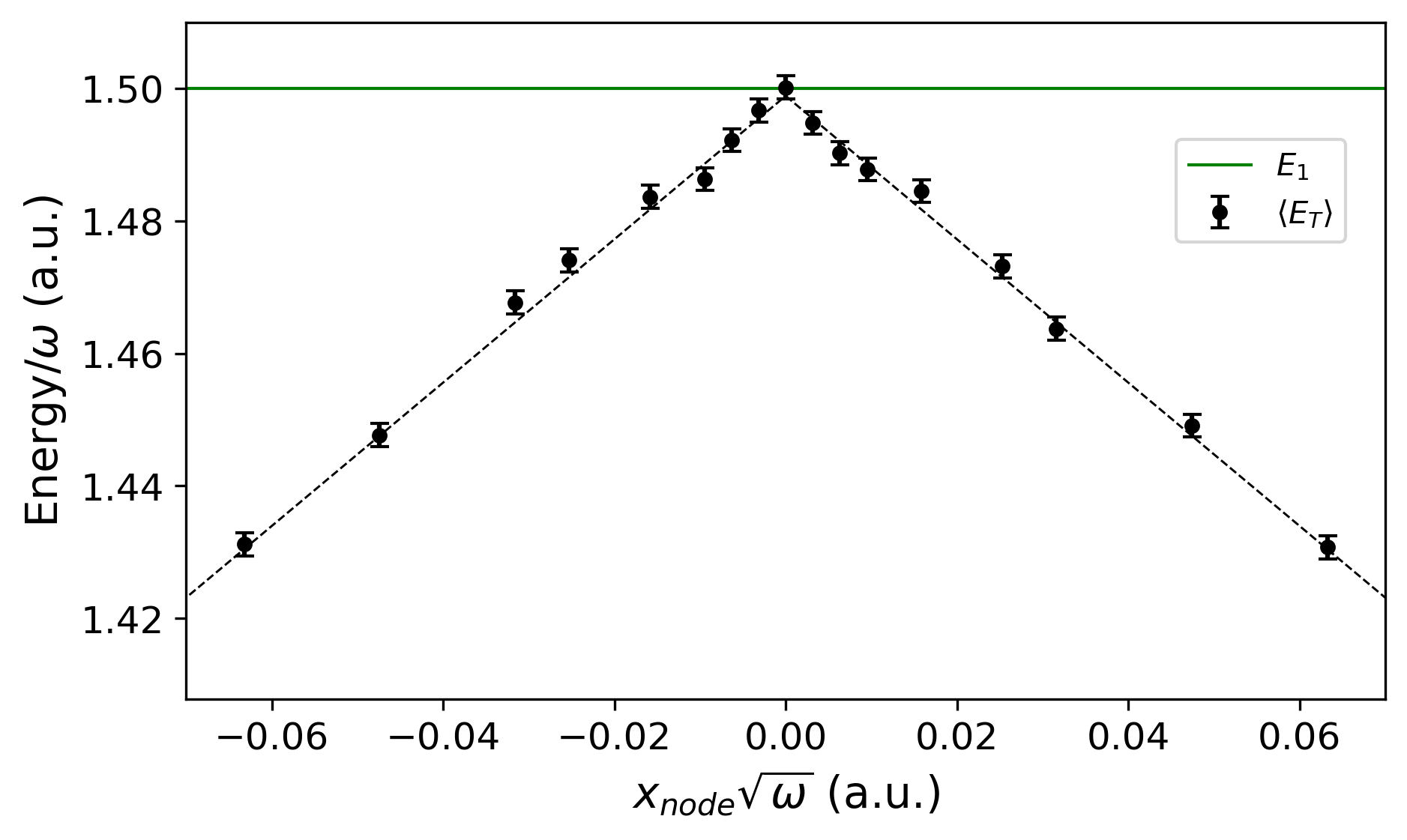}
    \caption{\small The error in the DMC energy of an excited state is first order in the node displacement. Due to inversion symmetry, the data obtained with $x_{node}>0$ are consistent within the error bars with the $x_{node}<0$ ones. Simulations performed with $\delta\tau=1.6\cdot 10^{-5}$ a.u., $\tau=410$ a.u.\ and $N_{\rm target}=10^4$.}
    \label{fig: harm1node node error T410}
    \end{center}
\end{figure}
For large node displacement, instead, we observe how the growth estimator stabilises around a minimum value equal to $E_0=0.5$ a.u.\ (Fig.~\ref{fig: harm1node node error T25610}).
\begin{figure}
    \begin{center}
    \includegraphics[width=0.68\textwidth]{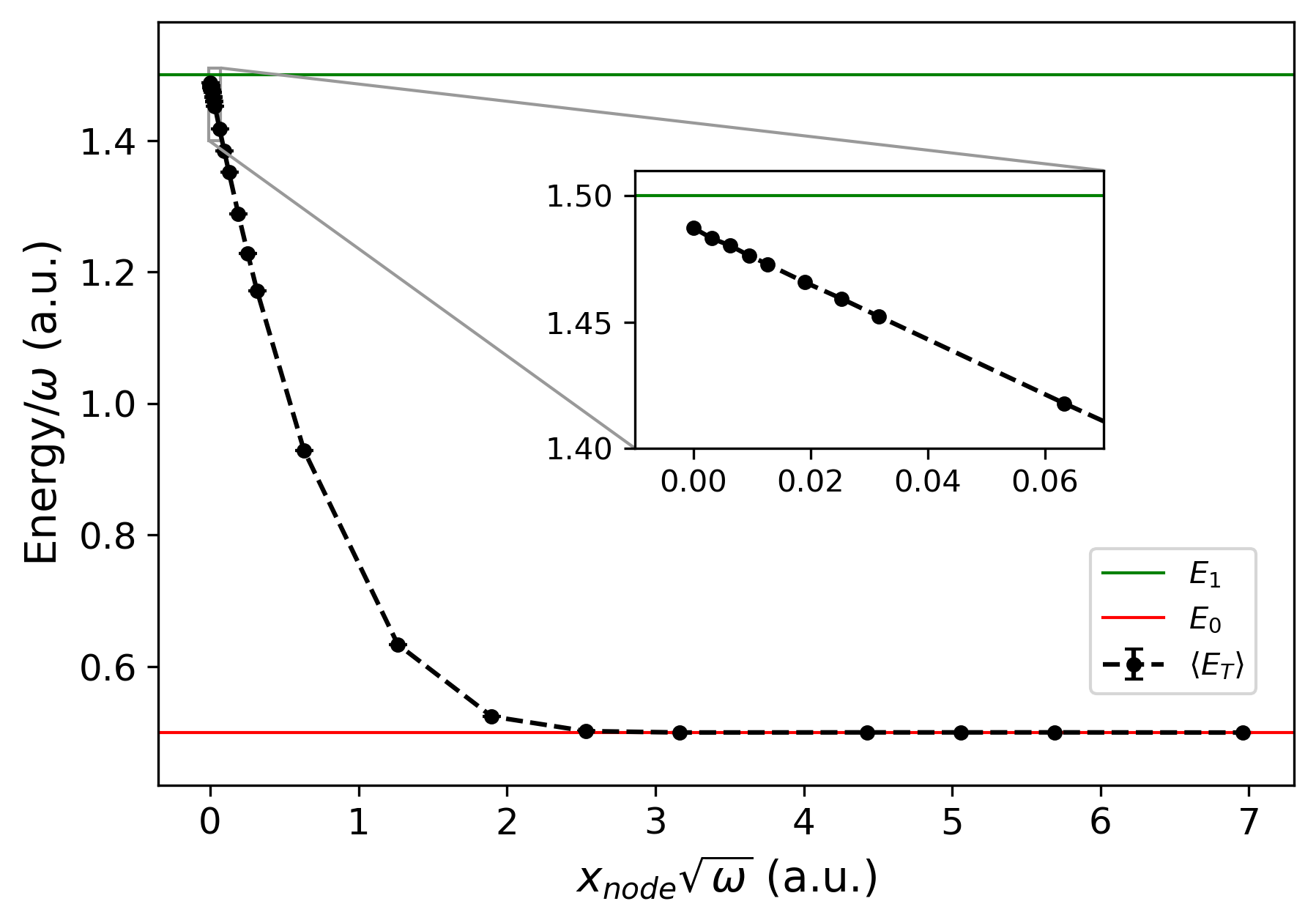}
    \caption{\small The fixed-node DMC energy is linear in the zero limit (inset window) and gradually converges to the ground state energy $E_0$ (red line) as a larger $x_{node}$ is enforced. The discrepancy with $E_1$ (green line) at $x_{node}=0$ is due to the time step bias, which does not affect the overall qualitative behaviour. Error bars are smaller than the size of the symbols. Simulations performed with $\delta\tau=0.001$ a.u., $\tau=25610$ a.u.\ and $N_{\rm target}=10^4$.}
    \label{fig: harm1node node error T25610}
    \end{center}
\end{figure}
Although the variational principle does not hold for the first excited state, i.e.\ $\langle E_{T} \rangle \,(x_{node})\leq E_1$, it is still valid for the ground state. Then large value of $x_{node}$ implies a large overlap between the walkers distribution and the nodeless ground state $\psi_0$ and an energy estimate close to (but never below) $E_0$, according to Eq.~\ref{eq: energy mixing}.

In addition, the absence of the tiling property leads to a strong dependence on the initial condition. If we generate all the walkers inside the unfavourable pocket, the dynamics is constrained there due to the fixed node and eventually the distribution cannot sample the lowest of the pocket eigenvalues. Accordingly, the populated region in Fig.~\ref{fig: harm1node +0.2 pocket+ phi}(b) is where $x>x_{node}$ and the energy estimate is higher than $E_1$. Its behaviour is still linear for small node displacement, but now with a positive slope for $x_{node}>0$ equal to $b_{>0}=1.10\pm 0.02$, as derived from data fitting in Fig.~\ref{fig: harm1node node error T410 pocket+}. Not surprisingly the (negative) slope from Fig.~\ref{fig: harm1node node error T410} is equal to $b_{<0}=-1.08 \pm 0.01$, sharing the same absolute value but different sign with $b_{>0}$ within the error.
\begin{figure}
    \begin{center}
    \includegraphics[width=11cm]{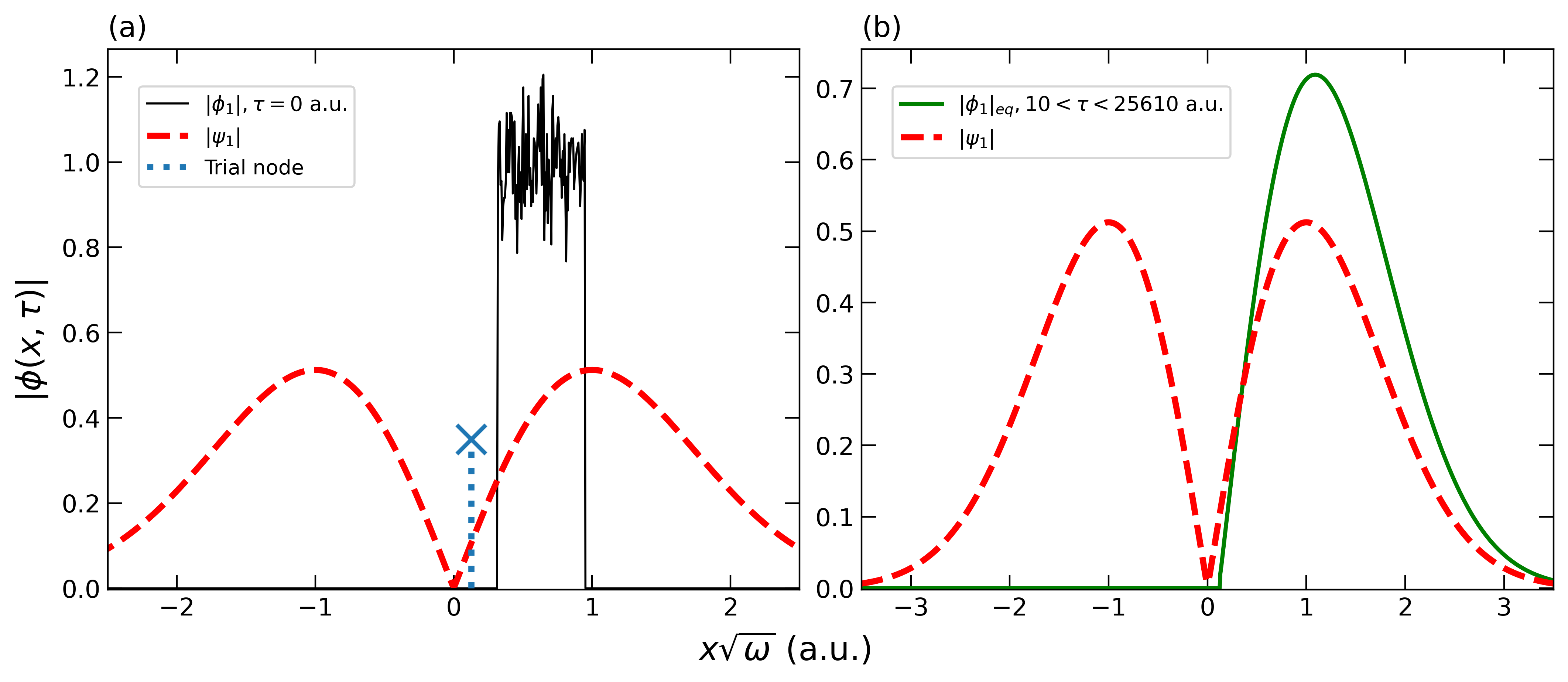}
    \caption{\small Fixing the node at $x_{node}>0$ and (a) populating just the unfavourable nodal pocket as initial condition lead to (b) a final (averaged) distribution of walkers entirely enclosed into the $\bigl\{x>x_{node}\bigr\}$ region and a DMC energy larger than $E_1$. $x_{node}\cdot \sqrt{\omega}=0.2\,\sqrt{0.4}\approx 0.13$ a.u., $\delta\tau=0.001$ a.u., $N_{\rm target}=10^4$.}
    \label{fig: harm1node +0.2 pocket+ phi}
    \end{center}
\end{figure}
\begin{figure}
    \begin{center}
    \includegraphics[width=0.78\textwidth]{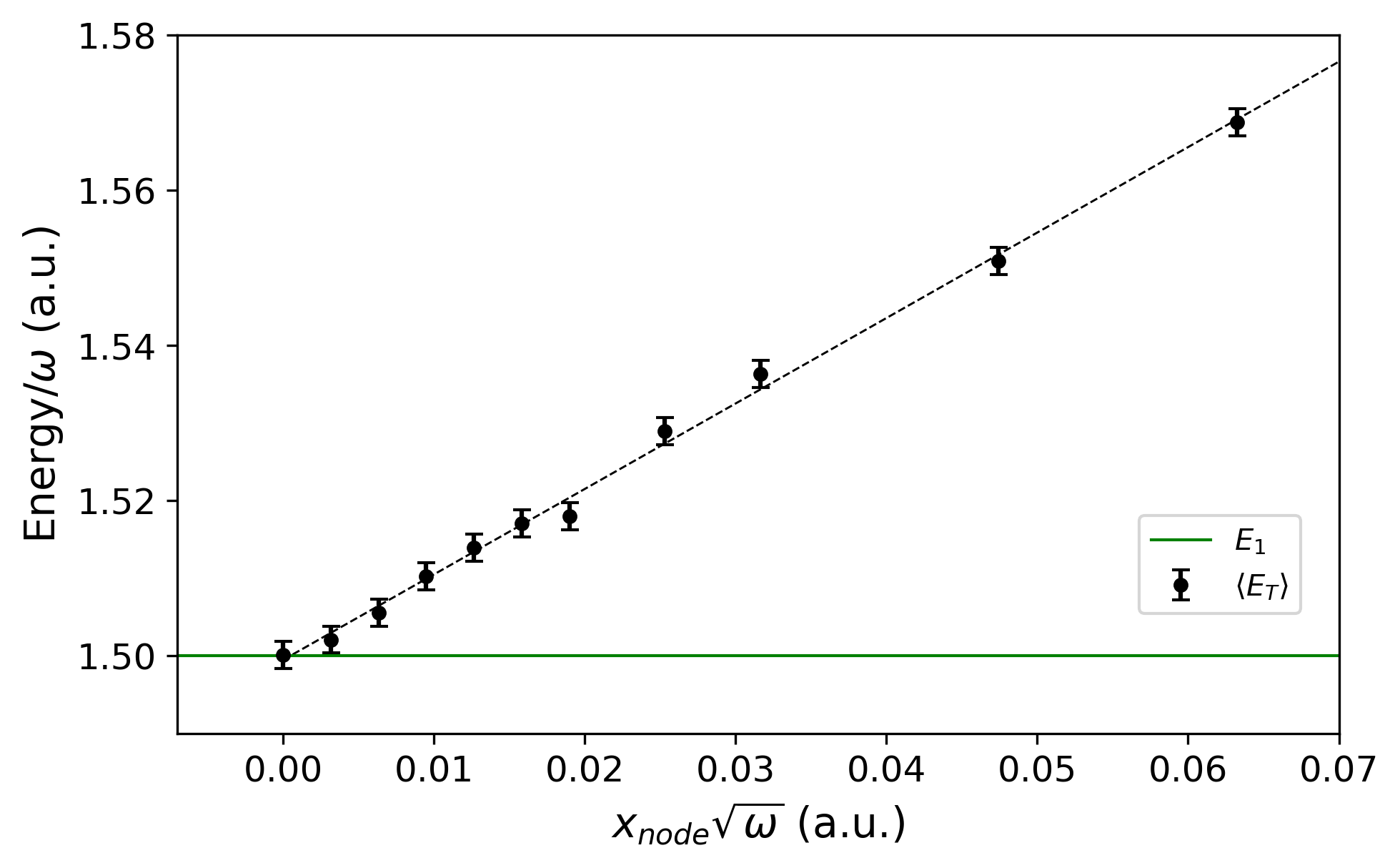}
    \caption{\small If the population samples only the unfavourable nodal pocket, the error in the fixed-node DMC energy is still first order in the node displacement, but now $\langle E_{T} \rangle \,(x_{node})\geq E_1$. $\delta\tau=1.6\cdot 10^{-5}$ a.u., $\tau=410$ a.u., $N_{\rm target}=10^4$.}
    \label{fig: harm1node node error T410 pocket+}
    \end{center}
\end{figure}
\\
For the excited state of a multi-particle system in two (or more) dimensions, the above analysis is not so simple. Even if its nodal surface is exactly known, which is not usually the case, it is difficult to parameterise in a systematic fashion and no intuitive representation is available. In any case, we expect the same features: the possible lack of variationality and a linear dependence of the DMC energy on the fixed-node error.

\section{Diffusion Monte Carlo with Importance sampling}\label{sec: importance sampling}
Throughout this tutorial, we have used, for the sake of clarity, the simplest DMC algorithm, which is easier to implement, {yet limited in capability,} and which nevertheless allows the main aspects of DMC to be grasped. However, production codes for the study of electronic structure of molecules, solids and surfaces employ a more efficient, but also more elaborate version of it, which ultimately becomes the best approach in all practical cases.
The procedure explained in the previous sections is completely general and do not require any prior knowledge of the ground state wavefunction of the Hamiltonian, however, if some knowledge exists, then this can be used to make the algorithm more efficient, increasing the sampling where the wavefunction is expected to be large \cite{PhysRevLett.45.566,10.1063/1.440575,10.1063/1.441022,10.1007/BFb0018166,10.1063/1.443766,SUHM1991293}. 
If knowledge of the ground state of the system is coded into a {\em trial wavefunction $\psi$}, this differential sampling is achieved by modifying the original algorithm as follows.

Recall the Schr\"odinger equation in imaginary time Eq.~\ref{eqn:2}, which we rewrite as: 
\begin{flalign}\label{eqn:32} 
 \frac{\partial \phi}{\partial \tau} = \frac{1}{2}\nabla^2 \phi + (E_T - V)\phi.
\end{flalign}
We rearrange it now in terms of the product $f(\tau) = \psi \phi(\tau)$. To do that, we need expressions for $\partial f/\partial \tau$ and the laplacian of $\phi$ in terms of the gradient and the laplacian of $f$. Since $\psi$ does not depend on time, we immediately have:
\begin{flalign}\label{eqn:33} 
 \frac{\partial f}{\partial \tau} = \psi \frac{\partial \phi}{\partial \tau}.
\end{flalign}
We then have:
\begin{flalign}\label{eqn:34} 
 \nabla \phi &= \frac{\nabla f}{\psi} -f \frac{\nabla \psi}{\psi^2} = \frac{1}{\psi}\left [ \nabla f - f \frac{\nabla \psi}{\psi} \right ], \nonumber \\
\nabla^2 \phi & = -\frac{\nabla \psi }{\psi^2}\cdot \left [ \nabla f - f \frac{\nabla \psi}{\psi} \right ] + \frac{1}{\psi}\nabla\left [ \nabla f - f \frac{\nabla \psi}{\psi} \right ], \nonumber \\
\psi \nabla^2 \phi &= -\frac{\nabla \psi}{\psi}\cdot \nabla f + f \left ( \frac{\nabla \psi}{\psi} \right )^2  + \nabla^2 f \nonumber\\
&- \nabla f \cdot \frac{\nabla \psi}{\psi} -f \frac{\nabla^2 \psi}{\psi} + f  \left ( \frac{\nabla \psi}{\psi} \right )^2  \nonumber \\
& = \nabla^2 f -f \frac{\nabla^2 \psi}{\psi} +2 f \left ( \frac{\nabla \psi}{\psi} \right )^2  -2 \frac{\nabla \psi}{\psi}\cdot \nabla f. 
\end{flalign}
Combinining Eqs.~\ref{eqn:32},~\ref{eqn:33} and~\ref{eqn:34} we obtain:
\begin{flalign}\label{eqn:35} 
\frac{\partial f}{\partial \tau} = \frac{1}{2}\nabla^2 f -\frac{1}{2}f \frac{\nabla^2 \psi}{\psi} +f \left ( \frac{\nabla \psi}{\psi} \right )^2  - \frac{\nabla \psi}{\psi} \cdot \nabla f + (E_T - V)f.
\end{flalign}
It is useful to write $-\frac{1}{2}f \frac{\nabla^2 \psi}{\psi}  = \frac{1}{2}f \frac{\nabla^2 \psi}{\psi} - f \frac{\nabla^2 \psi}{\psi}$, and  
define the {\em local energy}  $E_L({\bf R}) \equiv \hat{H}\psi({\bf R}) / \psi({\bf R})$, so that  Eq.~\ref{eqn:35} becomes:
\begin{flalign}\label{eqn:36} 
\frac{\partial f}{\partial \tau} = \frac{1}{2}\nabla^2 f -f \frac{\nabla^2 \psi}{\psi} +f \left ( \frac{\nabla \psi}{\psi} \right )^2  - \frac{\nabla \psi}{\psi} \cdot \nabla f + (E_T - E_L)f.
\end{flalign}
Finally, we define the {\em drift velocity} ${\bf v}({\bf R}) \equiv \nabla \psi({\bf R})/\psi({\bf R})$ (or {\em local gradient}), and noting:
\begin{flalign}\label{eqn:37} 
\nabla ({\bf v} f) = \nabla f \cdot \frac{\nabla \psi}{\psi} + f \frac{\nabla^2 \psi}{\psi} -  f \left ( \frac{\nabla \psi}{\psi} \right )^2,
\end{flalign}
Eq.~\ref{eqn:36} becomes:
\begin{flalign}\label{eqn:38} 
\frac{\partial f}{\partial \tau} = \frac{1}{2}\nabla^2 f - \nabla ({\bf v} f) + (E_T - E_L)f.
\end{flalign}
Eq.~\ref{eqn:38} is similar to Eq.~\ref{eqn:32}, but with two important differences. 
The first is that the potential energy $V$ is replaced by the local energy $E_L$, which for a good trial wavefunction $\psi$ can be much smoother. For instance, it could satisfy the {\em cusp condition} \cite{Kato,10.1063/1.1727605} on the divergence of the Coulomb potential and not diverge there.
In fact, a cusp in the exact wavefunction, i.e.\ a discontinuity in its first derivative, is required to cancel out the singularity of the potential with a corresponding term in the local kinetic energy.
Thus, in the limit of $\psi$ being an eigenstate $\psi_n$ of the Hamiltonian with eigenvalue $E_n$, $E_L({\bf R})$ would be equal to $E_n$ for every value of ${\bf R}$. Since $E_L({\bf R})$ is replacing $V({\bf R})$ in the branching factor of the Green function, $G_r({\bf R},{\bf R}^\prime,\delta \tau) = e^{ -\delta \tau \left ( \frac{E_L({\bf R}) + E_L({\bf R}^\prime)}{2} - E_T \right )}$, this results in much weaker fluctuations and a more stable population of walkers.

The second difference comes from the extra $\nabla ({\bf v} f)$ term, which is not present in Eq.~\ref{eqn:32}. To understand the effect of this term, it is useful to introduce an approximation and assume that for short enough time steps ${\bf v}$ is constant over an evolution of time $\delta \tau$, and so $\nabla ( {\bf v} f ) \approx {\bf v}\cdot \nabla f$. This allows us to write $ \frac{1}{2}\nabla^2 f - {\bf v}\cdot \nabla f = (-\hat{p}^2/2 - i {\bf v} \cdot \hat{p})f$, and so the diffusion part of the Green function can be written as:
\begin{flalign}\label{eqn:39} 
G_d({\bf R},{\bf R}^\prime,\delta \tau) = \langle {\bf R}| e^{ -\delta \tau \frac{\hat{p}^2}{2} - i \delta \tau {\bf v}\cdot  \hat{p}}|{\bf R}^\prime \rangle.
\end{flalign}
By inserting a resolution of the identity and working in a similar way as in Sec.~\ref{chap:shim}, we arrive at the following expression for a {\em drift-diffusion} Green function:
\begin{equation}\label{eqn:40} 
  G_d({\bf R},{\bf R^\prime},\tau)  = \left ( \frac{1}{2\pi\tau}\right )^\frac{3}{2}e^{-\frac{|{\bf R} - {\bf R}^\prime - \delta \tau {\bf v}|^2 }{2\tau}}.
\end{equation}
We see that in addition to the diffusion process there is a {\em drift} caused by the term ${\bf v} = \nabla\psi/\psi$. This quantity is large where $\psi$ is small and decreasing/increasing, for example near the nodal surface of the wavefunction. The $ \delta \tau {\bf v}$ term then pushes the walker away from the nodes and it is proportional to $1/R_\perp$ close to them, where $R_\perp$ is the distance in the direction normal to the nodal surface. In other words, the drift term, which diverges on the nodal surface, helps to keep the density of walkers low where $\psi$ is small.
Hence, the importance sampling technique provides a natural way to implement the fixed-node approximation, because in the limit of zero time step the multiplication by $\psi$ ensures that the steady state distribution has the same nodal surface of $\psi$.


The growth estimator $\langle E_T \rangle$ works also in the importance sampling formulation, whereas the one in Eq.~\ref{eqn:27} needs to be adapted by replacing the arbitrary function $\phi$ with the trial wavefunction $\psi$ and considering that $\ket{\psi_0}=\lim_{\tau \rightarrow \infty}\ket{e^{-\tau(\hat{H}-E_0)}\psi}$:
\begin{equation}
\begin{aligned}[b]
  E_0 & =\frac{\braket{\psi_0|\hat{H}|\psi_0}}{\braket{\psi_0|\psi_0}} \\ 
  & =\lim_{\tau \rightarrow \infty} \frac{\braket{e^{-2\tau(\hat{H}-E_0)}\psi|\hat{H}|\psi}}{\braket{e^{-2\tau(\hat{H}-E_0)}\psi|\psi}} \\
  & =\frac{\braket{\psi_0|\hat{H}|\psi}}{\braket{\psi_0|\psi}}.
  \label{eq: energy mixed estimator}
\end{aligned}
\end{equation}
Applying the local energy definition $E_L({\bf R}) = \hat{H}\psi({\bf R}) / \psi({\bf R})$ and recalling the walkers distribution $f(\mathbf{R},\tau)=\psi({\bf R})\phi(\mathbf{R},\tau)$, then:
\begin{equation}
\begin{aligned}[b]
  E_0 & \approx \lim_{\tau \rightarrow \infty} \frac{\int f(\mathbf{R},2\tau)E_L(\mathbf{R})\,d\mathbf{R}}{\int f(\mathbf{R},2\tau)\,d\mathbf{R}} = \langle E_L \rangle\\
  & \approx \frac{1}{n - n_{\rm equil}}\sum_{j=1}^{n-n_{\rm equil}} \frac{1}{N_j}\sum_{i=1}^{N_j}E_L({\bf R}_{ij}),
\end{aligned}
\end{equation}
where the average is computed over a simulation with $n$ steps, each of them evolving $N_j$ walkers at a time, and the associated stochastic error is:
\begin{flalign}\label{eq: importance sampling stochastic error} 
 \sigma_{\langle E_{L} \rangle}  = \sqrt{\frac{n_c}{n - n_{\rm equil}}}\sqrt{\langle E_{L}^2 \rangle - \langle E_{L} \rangle ^2}.
\end{flalign}
In literature this is known as the {\em mixed estimator} \cite{PhysRevE.51.3679}.
{Let's stress that the walkers population no longer samples the ground state wavefunction, but rather the \emph{mixed} distribution $f(\mathbf{R},\tau\rightarrow\infty)=\psi({\bf R})\psi_0(\mathbf{R})$. As a result, the mixed estimator is inherently biased by the chosen trial wavefunction $\psi$.}
{From Eq.~\ref{eq: energy mixed estimator} it is straightforward to verify that the mixed estimator can compute exact values in the same way for every operator that commute with the Hamiltonian, e.g.\ every function of the Hamiltonian operator, such as atomization energy, ionization potential, and electron affinity, simply obtained as a difference of energies. However, for every other operator $\hat{A}$ (for example, the electric dipole operator) the mixed estimator does not yield the correct expectation value $A_0$, i.e.:
\begin{equation}
    A_0\equiv \frac{\braket{\psi_0|\hat{A}|\psi_0}}{\braket{\psi_0|\psi_0}}\neq\frac{\braket{\psi_0|\hat{A}|\psi}}{\braket{\psi_0|\psi}}.
\end{equation}
Therefore, both exact and approximated methods to sample from the “pure" $\psi_0^2$ distribution have been developed. In that way the expectation value $A_0$ becomes accessible from QMC calculations. For example, it can be proved that, while both the mixed estimator $A_m$ and the \emph{variational estimator}:
\begin{equation}
    A_v\equiv \frac{\braket{\psi|\hat{A}|\psi}}{\braket{\psi|\psi}}
\end{equation}
poorly approximate $A_0$, they can be combined to provide a better approximation, namely:
\begin{equation}
    2A_m-A_v = A_0 + O (\Delta^2),
\end{equation}
where $\Delta\equiv \psi_0- \psi$ \cite{QMCbook_Hammond,Ceperley1986}.}

The challenges in implementing this enhanced algorithm from scratch consist mainly in constructing and optimising a trial wavefunction specific to each quantum system. In general, a good trial wavefunction is required to strike a balance between accuracy and ease of evaluation. A compact and widespread representation typically employ \emph{Slater-Jastrow} (SJ) functions. A SJ trial wavefunction consists of a single Slater determinant multiplied by a symmetric non-negative function of the electronic coordinates, which is dubbed Jastrow factor. The Jastrow factor accounts for the electronic correlation and satisfy the aforementioned electron-electron and electron-nuclei \emph{cusp conditions}. The orbitals within the Slater determinant can be derived from precise DFT or Hartree-Fock (HF) calculations, or even from more sophisticated approaches depending on the required level of accuracy \cite{10.1063/1.3697846}.
The Jastrow factor can have several functional forms and crucially depends on the optimization of its parameters. For a $N$ particles system, a SJ wavefunction can be written as:
\begin{equation}
  \psi_{SJ}(\mathbf{X})=e^{J(\mathbf{X})}D(\mathbf{X}),
  \label{eq: Slater-Jastrow wavefunction}
\end{equation}
where $D(\mathbf{X})$ is the Slater determinant, $\mathbf{X}=(\mathbf{x}_1,\mathbf{x}_2,...,\mathbf{x}_N)$ and $\mathbf{x}_i=\{\mathbf{r}_i,\sigma_i\}$, i.e.\ spatial and spin coordinates. Usually the Jastrow factor has the following form:
\begin{equation}
  J(\mathbf{X})=\sum_{i=1}^N\chi(\mathbf{x}_i)-\frac{1}{2}\sum_{i\neq j}^N u(\mathbf{x}_i,\mathbf{x}_j).
\end{equation}
Here the $\chi$ terms describe the electron-nuclei correlation, while the $u$ terms describe the electron-electron correlation. 
Depending on the investigated system, those functions can be defined in several ways and even three-body correlation terms could be included. 
One of the most diffused form for the $u$ function in the Jastrow factor is the following:
\begin{equation}
  u(\mathbf{x}_i,\mathbf{x}_j)=\frac{A_{\sigma_i,\sigma_j}}{r_{ij}}(1-e^{-r_{ij}/B_{\sigma_i,\sigma_j}}),
\end{equation}
which was used in some of the earliest work on the homogeneous electron gas \cite{PhysRevLett.45.566}. Here $r_{ij}=|\mathbf{r}_{ij}|=|\mathbf{r}_{i}-\mathbf{r}_{j}|$ and the parameters $A_{\sigma_i,\sigma_j}$ and $B_{\sigma_i,\sigma_j}$ are fixed by imposing both the cusp conditions and the proper long-range behaviour.

\section{Further reading}
In this last section, we aim to supply the reader with a comprehensive list of sources to delve into the more advanced features of the QMC methods and eventually approach the recent literature.

A general overview, with a focus on the theoretical aspects of the diffusion Monte Carlo, is provided by Foulkes {\em et al.}'s review \cite{Foulkes2001}, which we have referred to extensively during the production of this tutorial. The books by Hammond {\em et al.} \cite{QMCbook_Hammond} and Becca and Sorella \cite{becca_sorella_2017} also introduce the fundamentals of statistics and Monte Carlo sampling (e.g.\ Markov chain theory and the Metropolis algorithm) as essential tools for the reader. Great attention is paid to the different types of Quantum Monte Carlo methods other than the diffusion one, such as the variational method or the exact Green's function method.

The practical implementation of QMC as well as all aspects related to the efficiency, reliability and limitations of the algorithm are usually presented in the papers or guides supporting recent versions of QMC production codes. Among the most popular ones, it is worth mentioning CASINO \cite{casino, 10.1063/1.5144288}, TurboRVB \cite{10.1063/5.0005037}, QMCPACK \cite{Kim_2018}, and CHAMP \cite{filippi_champ}.
As already stated, they all employ the importance sampling together with the enhanced version of the algorithm from Umrigar {\em et al.} \cite{10.1063/1.465195} in order to achieve faster and more accurate calculations.
Since trial wavefunctions and their nodal surfaces play a significant role in simulations of fermionic systems, they still represent an active research topic nowadays, with a main focus on their optimisation. We recommend the review by Austin, Zubarev and Lester \cite{doi:10.1021/cr2001564} as it offers an insightful overview on this field.

Coming to the applications, QMC has made significant progress in several areas of electronic structure theory thanks to its favorable properties, especially in those where electron correlation is a key or dominant factor. 
QMC accuracy is suitable particularly for systems with weak intermolecular interactions, where a subtle balance in bonding arises from long-range dynamical correlations, i.e.\ dispersion effects. A comprehensive review on non-covalent interactions addressed by Quantum Monte Carlo methods is that of Dubecký {\em et al.} \cite{doi:10.1021/acs.chemrev.5b00577}.
Furthermore, QMC is increasingly employed in conjunction with quantum chemistry methods to provide systematic benchmark quality results for molecules. A nice discussion on this topic can be found in Al-Hamdani and Tkatchenko \cite{10.1063/1.5075487}, where the DMC is regarded as a worthy competitor to the Coupled Cluster with single, double, and perturbative triple [CCSD(T)] excitations, often considered the “gold standard” for many chemistry applications.

\section*{Conclusions}
\addcontentsline{toc}{section}{Conclusions}
In this tutorial, we have provided an essential explanation of the diffusion Monte Carlo method, designed to simultaneously calculate the ground state energy and wavefunction of any quantum system. We developed a straightforward numerical algorithm and implemented a computer program to determine the ground state of a few simple systems of educational interest.
The codes, inputs and outputs, of the examples illustrated here are available on {\href{https://github.com/zenandrea/QMC-tutorial}{GitHub}}. In this tutorial we both describe the simple and importance sampling version of the DMC algorithm. For seek of simplicity, in the illustrated examples we employed the simple algorithm, which is less efficient than the importance sampling algorithm, but way easier to implement. 

The typical systematic errors affecting a DMC calculation, i.e.\ the time step bias and the finite population bias, have been studied and the extrapolation to $\delta\tau\rightarrow0$, $N_{\rm target}\rightarrow\infty$ returned accurate energy results (e.g.\ the statistical error on energies in Table \ref{tab: fit parameters} is below a thousandth of $\omega$ and could be further improved with longer runs). 
The hydrogen atom turned out to be computationally demanding due to its three-dimensionality and the presence of a singularity in the Coulomb potential, which has been taken under control by imposing a threshold on the walker replication.
The problem of instability in DMC simulations is mitigated by the importance sampling algorithm, but not completely removed, so a control of instability by limiting the population size is a general feature of all DMC implementations \cite{10.1063/1.465195,10.1063/1.4917171}.
In order to apply the DMC method to systems of interacting fermions one has to treat, as mentioned above, the “sign problem” due to the antisymmetry property of the many-fermion wavefunction. Here we have implemented the fixed-node approximation by the killing procedure, that is walkers crossing the nodal surface are deleted.
It has been shown that the time step error follows a power law, and that this is quadratic for the nodeless ground state, but only proportional to $\delta \tau^{1/2}$ within our algorithm for any higher-energy state (including fermionic ones).

We also artificially imposed trial node errors and showed the implications. By slightly moving the trial nodal surface away from its exact position, we have shown that the energy increases quadratically in ground state calculations, while it may decrease linearly and exhibit a slope discontinuity when dealing with excited states.
This last situation arises from the violation of the variational principle and the tiling property which may occur depending on the chosen trial nodal surface.
Indeed, for the first excited state of the harmonic oscillator, the exact node will be fixed completely by the inversion symmetry of the Hamiltonian. In more complex problems, symmetry arguments do not define the exact topology of the nodal surface, but can constrain our guess and even restore the variational principle when imposing orthogonality to every lower-energy state.

In summary, DMC is a versatile and accurate method for simulations of quantum systems, ranging from atoms \cite{10.1063/1.3554625}, molecules and clusters \cite{10.1063/1.4730035} to condensed matter \cite{Kolorenč_2011}.
Despite not directly addressing real systems in this work, we take a step towards democratising DMC and lowering the learning barrier by providing a beginner-friendly discussion of both the physics and the algorithmic complexities behind this powerful method.

\section*{Supplementary Material}
See {\href{https://github.com/zenandrea/QMC-tutorial}{supplementary material}} for the uniform sampling DMC codes and the Jupyter Notebook files to generate the figures in the article.

\section*{Acknowledgments}
A.A., D.A., and A.Z. acknowledge support from the European Union under the Next generation EU (projects 20222FXZ33 and P2022MC742).
D.A. and A.Z. also acknowledge support from Leverhulme grant no. RPG-2020-038.
The authors acknowledge the use of the IBISCO cluster (funded by project code PIR01\_00011 “IBISCo”, PON 2014-2020).

\bibliographystyle{unsrt}
\bibliography{bibliography}

\end{document}